\documentclass[11pt]{article}

\usepackage{geometry}  
\usepackage[english]{babel}
\usepackage[utf8]{inputenc}
\usepackage[T1]{fontenc}

\usepackage{xr}

\usepackage[dvipsnames]{xcolor}
\usepackage{paralist}
\usepackage{graphicx}
\usepackage{subcaption}
\usepackage{longtable} 
\usepackage{multirow}
\usepackage{listings}
\usepackage{makecell}
\usepackage{array}
\usepackage{float}
\usepackage{dsfont}
\usepackage{rotating}
\usepackage{booktabs}
\usepackage{enumerate}
\usepackage{tikz}
\usepackage{pgf}
\usepackage{enumitem}
\newfloat{algorithm}{t}{lop}
\usepackage{lipsum} 
\usepackage{titlesec}

\usepackage{amsmath,amssymb,amsfonts,amsthm,bbm}
\usepackage{mathtools}
\usepackage{mathrsfs}
\usepackage{algorithm}
\usepackage{algpseudocode}
\usepackage{algorithmicx}

\usepackage{natbib}
\usepackage{authblk}
\usepackage{todonotes}
\usepackage{xr-hyper}
\usepackage[
  colorlinks,
  citecolor=blue,
  linkcolor=red,
  anchorcolor=red,
  urlcolor=blue
]{hyperref}

\usepackage{tikz}
\usetikzlibrary{shapes.geometric, arrows}
\tikzstyle{startstop} = [rectangle, rounded corners, minimum width=3cm, minimum height=1cm, text width=3cm, text centered, draw=black, fill=red!30]

\def\orderless{\preceq}
\allowdisplaybreaks

\usepackage{color}
\definecolor{cm}{RGB}{200,0,0}

\definecolor{rfb}{RGB}{0,150,150}

\definecolor{yg}{RGB}{138,43,226}

\def\iso{\mathrm{iso}}

\def\la{\langle}
\def\ra{\rangle}

\newlist{todolist}{itemize}{2}
\setlist[todolist]{label=$\square$}
\usepackage{pifont}
\usepackage{mdframed}
\usepackage{kantlipsum}

\mdfdefinestyle{myenvs}{%
  hidealllines=true,%
  nobreak=true, 
  leftmargin=0pt,
  rightmargin=0pt,
  innerleftmargin=0pt,
  innerrightmargin=0pt,
}

\theoremstyle{plain}
\newmdtheoremenv[style=myenvs]{theorem}{Theorem}[section]
\newmdtheoremenv[style=myenvs]{proposition}[theorem]{Proposition}
\newmdtheoremenv[style=myenvs]{lemma}[theorem]{Lemma}
\newmdtheoremenv[style=myenvs]{example}[theorem]{Example}
\newmdtheoremenv[style=myenvs]{corollary}[theorem]{Corollary}
\theoremstyle{definition}
\newmdtheoremenv[style=myenvs]{definition}[theorem]{Definition}
\newmdtheoremenv[style=myenvs]{assumption}[theorem]{Assumption}
\newtheorem{condition}[theorem]{Condition}
\theoremstyle{remark}

\newcommand{\ind}{\mathbbm{1}}
\newcommand{\var}{\mathrm{Var}}

\newcommand{\argmin}[1]{\underset{#1}{\arg\!\min}}

\usepackage{xspace}

\def\##1\#{\begin{align}#1\end{align}}
\def\$#1\${\begin{align*}#1\end{align*}}

\definecolor{myblue}{rgb}{.8, .8, 1}
\definecolor{mathblue}{rgb}{0.2472, 0.24, 0.6} 
\definecolor{mathred}{rgb}{0.6, 0.24, 0.442893}
\definecolor{mathyellow}{rgb}{0.6, 0.547014, 0.24}

\newcommand{\calB}{{\mathcal{B}}}
\newcommand{\calC}{{\mathcal{C}}}
\newcommand{\calD}{{\mathcal{D}}}

\newcommand{\calG}{{\mathcal{G}}}

\newcommand{\calN}{{\mathcal{N}}}

\newcommand{\calP}{{\mathcal{P}}}
\newcommand{\calQ}{{\mathcal{Q}}}
\newcommand{\calR}{{\mathcal{R}}}

\newcommand{\calV}{{\mathcal{V}}}

\newcommand{\calX}{{\mathcal{X}}}
\newcommand{\calY}{{\mathcal{Y}}}

\newcommand{\bfm}[1]{\ensuremath{\mathbf{#1}}}

  \def\EE{\mathbb{E}}

 \def\bI{\bfm I}

  \def\NN{\mathbb{N}}
  
  \def\PP{\mathbb{P}}
  
  \def\RR{\mathbb{R}}

\def\vareps{\varepsilon}

\def\hat{\widehat}
\def\tilde{\widetilde}
\def\tr{\mathrm{train}}

\def\te{\mathrm{test}}

\DeclareTextFontCommand{\texttt}{\ttfamily\upshape}

\def\Du{\Delta(R;\calB)}
\def\Dul{\Delta_2(R;\calB)}

\def\Diso{\Delta^{\iso}(R;\calB)}
\def\hatDiso{\hat \Delta^{\iso}(R;\calB)}

\def\DRiso{\Delta(\pi(R);\calB)}
\def\hatDisor{\hat\Delta^{\iso}(r;\calB)}

\def\hatDriso{\hat\Delta(r^{\iso};\calB)}
\def\hatDRiso{\hat\Delta(R^{\iso};\calB)}

\def\hatDisoU{\hat \Delta^{\iso}(\calB)}

\def\Df{\Delta_{\rho}(R;\calB_{f,\rho})}

\def\Dfiso{\Delta^{\iso}(R;\calB_{f,\rho})}

\def\DfisoG{\Delta^{\iso}(R;\calB_{f,\rho,\wbnd})}

\def\Dbnd{\Delta(R;\calB_{a,b})}

\def\Dbndiso{\Delta^{\iso}(R;\calB_{a,b})}

\def\hatDbndr{\hat \Delta(r;\calB_{a,b})}

\def\hatDbndisor{\hat \Delta^{\iso}(r;\calB_{a,b})}
\def\hatDbndisoU{\hat \Delta^{\iso}(\calB_{a,b})}

\def\Disow{\Delta^{\iso}(R;\calB,\wprior)}
\def\tDisow{\tilde \Delta^{\iso}(R;\calB,\wprior)}

\def\Dw{\Delta(R;\calB,\wprior)}

\def\altDwiso{\Delta(\pi_1(\tilde R) \circ \wprior;\calB,\wprior)}

\def\wfiso{w^{*\iso}_{f,\rho}}

\def\hatDcal{\hat \Delta^{\iso}(\alpha)}
\def\myhash{\char"0023}
\def\Ptar{P_{\mathrm{target}}}
\def\tPtar{\tilde{P}_{\mathrm{target}}}
\def\tPtrain{\tilde{P}}
\def\Ptrain{P}

\def\wprior{w_0}

\def\alphaP{\EE_{\Ptrain}[R(X)]}

\def\Ciso{\calC^{\iso}_{\orderless}}
\def\Cisow{\calC^{\iso}_{\wprior}}
\def\CisoB{\calC^{\iso}_{\orderless,\wbnd}}

\def\Cisor{\calC^{\iso}_1}

\def\wbnd{\Omega}

\title{\mbox{Distributionally robust risk evaluation with an isotonic constraint}}
\author{Yu Gui}
\author{Rina Foygel Barber}
\author{Cong Ma}
\affil{Department of Statistics, University of Chicago}
\date{\today}

\begin{document}

\maketitle

\begin{abstract}
Statistical learning under distribution shift is challenging when neither prior knowledge nor data from the target distribution is available. Distributionally robust learning (DRL) aims to control the worst-case statistical performance within a set of candidate distributions, but how to properly specify the set remains challenging. To enable distributional robustness without being overly conservative, in this paper we propose a \mbox{shape-constrained} approach to DRL, which incorporates prior information about the way in which the unknown target distribution differs from its estimate---specifically, we assume the unknown density ratio between the target distribution and its estimate is isotonic with respect to some partial order. At the population level, we provide a solution to the \mbox{shape-constrained} optimization problem that can be solved without the challenge of an explicit isotonic constraint. At the sample level, we provide consistency results for an empirical
estimator of the target in a range of different settings. Empirical studies on both synthetic and real data demonstrate
the improved efficiency
of the proposed \mbox{shape-constrained} approach.
\end{abstract}



\section{Introduction}\label{sec:intro}
Evaluating the performance of an estimator is of significant importance in statistics. 
To give several motivating examples, we first consider 
supervised learning settings, where our observations consist of features
$X\in\calX \subseteq \RR^d$ and a response $Y\in \calY \subseteq \RR$:
\begin{itemize}
\item Given a fitted model $\hat\mu:\calX\to \RR$, we may want to estimate the expected value of the squared error $(Y-\hat\mu(X))^2$ with respect
to a target distribution on $(X,Y)$.
\item Or, in predictive inference, suppose we have constructed a prediction band $\hat{C}_{1-\alpha}$, where $\hat{C}_{1-\alpha}(X)\subseteq\RR$ is a confidence region for the response
$Y$ given features $X$, and $1-\alpha$ denotes the target coverage level. 
Then to determine whether $\hat{C}_{1-\alpha}$ does in fact achieve coverage at level $1-\alpha$
for data points drawn from some target distribution, we would like
to estimate the expected value of $\ind\{Y\not\in\hat C_{1-\alpha}(X)\}$ with respect
to this target distribution. This is the probability that our interval \emph{fails} to cover the response. 
\end{itemize}
We can also consider 
unsupervised learning settings, where observations consist only of features \mbox{$X\in\calX \subseteq \RR^d$}:
\begin{itemize}
\item In principal component analysis (PCA), suppose we have obtained a set of pre-fitted principal components $\widehat{\calV}_K = \{\hat{v}_1,\dots, \hat{v}_K\}$ which forms an orthonormal basis for a $K$-dimensional subspace of $\RR^d$. To evaluate how well the variance in $X$ is explained by the top $K$ principal components, it would be of interest to analyze the expected value of the reconstruction error $\| X - \sum_{k=1}^K (X^\top \hat{v}_k) \hat{v}_k \|^2$ with respect to the distribution of $X$.
\item Another example is density estimation. In this case, given a density estimate $P_{\theta}$ learned from data,  we may want to evaluate its performance using the expected log-likelihood  $-\log \mathsf{d}P_{\theta}(X)$ over a target distribution $\Ptar$. In fact, $\EE_{\Ptar}[-\log \mathsf{d}P_{\theta}(X)]$ is the cross-entropy of $P_{\theta}$ relative to $\Ptar$.
\end{itemize}
A key challenge for any of these problems is that the target distribution (say, the distribution
of the general population) may be unknown, and our available data (say, individuals who participate in our study) may be drawn
from a different distribution.

\subsection{Problem formulation}\label{sec:prob}
To make the problem more concrete, and unify the examples mentioned above, here we introduce some notation
to formulate the question at hand.

\paragraph{The unsupervised setting.}  Let $R:\calX\to\RR_+$ denote a \emph{risk function}, where our goal is to evaluate the expected value 
$\EE_{\Ptar}[R(X)]$ with respect to some target distribution $\Ptar$ over $\calX$. However, the available data
only provides information about $\Ptrain$, a potentially different distribution. 
Using a calibration data set comprised of samples $X_1,\dots,X_n$ drawn from $\Ptrain$,  we can estimate
$\EE_{\Ptrain}[R(X)]$ with the empirical mean, $n^{-1}\sum_{i=1}^n R(X_i)$). 
Our aim, though, is to provide a bound on the risk $\EE_{\Ptar}[R(X)]$---or, at least, to bound the difference in risks (often
called the \emph{excess risk}),
$\EE_{\Ptar}\left[R(X)\right] - \EE_{\Ptrain}\left[R(X)\right]$.

If we assume that the unknown distribution $\Ptar$ lies in some class $\calQ$ (to be specified later on),
then defining the \emph{worst-case excess risk}
\begin{align}\label{eq:opt-0}
\Delta(R;\calQ) = \sup_{Q \in \calQ}\EE_Q\left[R(X)\right] - \EE_{\Ptrain}\left[R(X)\right],
\end{align}
we can bound the risk under distribution~$\Ptar$ by $\EE_{\Ptar}\left[R(X)\right] \leq \EE_{\Ptrain}\left[R(X)\right] + \Delta(R;\calQ)$.

\paragraph{The supervised setting: covariate shift assumption.}
In the supervised learning setting, the data contains both features $X$ and a response $Y$. Here we will consider a loss function $r:\calX\times\calY\to \RR_+$, for instance, $r(x,y) = (y - \hat\mu(x))^2$ for the squared error in a regression, or $r(x,y) = \ind\{y\not\in\hat C_{1-\alpha}(x)\}$ for characterizing the (mis)coverage of a prediction interval in predictive inference.

Throughout this paper, for the supervised learning setting,
we will assume the \emph{covariate shift} setting, where the distribution of the available data and the target distribution
may differ in the marginal distribution of the covariates $X$, but share the same conditional distribution $Y\mid X$.
To make this concrete, if our calibration data consists of $n$ data points $(X_1,Y_1),\dots,(X_n,Y_n)$
drawn from $\tPtrain$, while our goal is to control the expected loss with respect to the target distribution $\tPtar$ on $(X,Y)$, we will assume
that we can write

\begin{align*}
\textnormal{data distribution: } \tPtrain = \Ptrain \times P_{Y\mid X},\quad 
\textnormal{target distribution: } \tPtar = \Ptar \times P_{Y\mid X},\end{align*}
so that $\tPtrain$ and $\tPtar$ share the same conditional distribution $P_{Y\mid X}$ for $Y\mid X$.

In fact, under covariate shift, this supervised setting can be unified with the unsupervised one by defining the risk $R(X) = \EE[ r(X,Y) \mid X]$,
which is the conditional expectation  of $r(X,Y)$ under \emph{either} $\tPtrain$ or $\tPtar$. 
The quantity of interest is then given by $\EE_{\Ptar}[R(X)] = \EE_{\tPtar}[r(X,Y)]$, 
but our calibration data, which is sampled from $P$, instead enables us to estimate $\EE_{\Ptrain}[R(X)]  = \EE_{\tPtrain}[r(X,Y)]$.
If we again assume that $\Ptar\in\calQ$,
then $\Delta(R;\calQ)$ again allows us
to bound the risk of our estimator under the target distribution:
\[\EE_{\tPtar}\left[r(X,Y)\right] \leq \EE_{\tPtrain}\left[r(X,Y)\right] + \Delta(R;\calQ).\]

\paragraph{Estimating the risk or tuning the model?}
In this paper, we consider the setting where our estimator---say, a prediction band $\hat C_{1-\alpha}$---is \emph{pretrained}, meaning
that we have available calibration data sampled from $\Ptrain$ (in the unsupervised setting) or $\tPtrain$ (in the supervised setting)
that is independent of the fitted estimator.
Consequently, our available calibration data provides us with an unbiased estimate of $\EE_{\Ptrain}[R(X)]$ (or, equivalently in the supervised
setting, $\EE_{\tPtrain}[r(X,Y)]$); given a constraint set $\calQ$, we can then use this estimate to bound $\EE_{\Ptar}[R(X)]$ (or,
in the supervised setting, $\EE_{\tPtar}[r(X,Y)]$).

In some settings, the goal may be to estimate the risk of each estimator within a family of (pretrained) options, in order to select a good estimator. Returning again to the example of a prediction band, suppose, we actually are given a nested family of prediction bands,
$\{\hat C_{1-a} : a\in[0,1]\}$, where $1-a$ denotes the confidence level.
Choosing $R_a(X) = \PP_{P_{Y|X}}(Y\not\in \hat C_{1-a}(X))$ or accordingly, $r_a(X,Y) = \ind\{Y\not\in\hat C_{1-a}(X)\}$, 
then, if we can compute a bound on the miscoverage rate $\EE_{\Ptar}[R_a(X)]$
for each $a$, then we can choose a value of $a$ that achieves some desired level of coverage.
More generally, we may do the same in other settings as well---that is, given a family of candidate estimators, bounding
 the risk of each one under the distribution $\Ptar$ provides an intermediate step towards choosing the tuning parameter.

Throughout this paper, then, we will primarily discuss the question of estimating the expected risk. Later on, in our experiments,
we will turn to the aim of using these estimates to tune a procedure for achieving a desired bound on the error.

\subsection{Prior work: distributionally robust learning} 
Our work builds upon the
distributionally robust learning (DRL) literature \citep{ben1998robust,el1998robust,lam2016robust,duchi2018learning},
which is a well-established framework for risk evaluation under distribution shift. 
In this framework, the target distribution $\Ptar$ is assumed to lie in some neighborhood around the distribution $\Ptrain$ of the available data---for instance,
we might assume that $ D_{\rm KL}(\Ptar\| \Ptrain)\leq \rho$, where $D_{\rm KL}$ denotes the  Kullbeck--Leibler (KL)  divergence.
DRL takes a conservative approach and evaluate the performance on $\Ptar$ via its upper bound, i.e., the worst-case performance over all distributions
within the specified neighborhood of $\Ptrain$,
\begin{equation}\label{eq:DRL}\EE_{\Ptar}[R(X)] \leq \sup\left\{\EE_Q[R(X)] : D_{\rm KL}(Q\| \Ptrain)\leq \rho\right\},\end{equation}
or, equivalently, $\EE_{\Ptar}[R(X)] \leq \EE_{\Ptrain}[R(X)] + \Delta(R;\calQ_{\rm KL}(\rho))$,
where $\Delta(R;\calQ_{\rm KL}(\rho))$ is defined as in~\eqref{eq:opt-0} with $\calQ = \calQ_{\rm KL}(\rho) =  \{ Q : D_{\rm KL}(Q\|\Ptrain)\leq \rho\}$.
More generally, we can consider divergence measures beyond the KL distance, as we will describe in more detail below.

\subsection{Our proposal: iso-DRL}\label{sec:intro-iso}
If the assumption $D_{\rm KL}(\Ptar\|\Ptrain)\leq \rho$ is correct,
then the upper bound~\eqref{eq:DRL} is valid.
However, since this bound uses only the KL divergence to define the constraint $\Ptar\in\calQ$ on the target distribution, 
it could be quite conservative. 
In many practical settings, additional side information or prior knowledge on the structure of the distribution shift may allow for a tighter
bound, which would be less conservative than the worst-case excess risk of DRL~\eqref{eq:DRL}. 
This raises the following key question:
\begin{center}
{\it Can we use side information on the distribution shift between the distribution $\Ptrain$ and the target distribution $\Ptar$, to improve the worst-case excess risk of DRL in risk evaluation?}
\end{center}
In this paper, we study one specific example of this type of setting: we
assume that the density ratio $(\mathsf{d}\Ptar/\mathsf{d}\Ptrain)(\cdot)$ between the target distribution and the data distribution
is (approximately) isotonic (i.e., monotone) with respect to some order or partial order on $\calX$.

\paragraph{Motivation: recalibration of an estimated density ratio.}

To motivate the use of such side information, consider a practical supervised setting where we have an initial estimate $\wprior$ for the density ratio:
\[\wprior(x)\approx \frac{\mathsf{d}\Ptar}{\mathsf{d}\Ptrain}(x).\]
This ratio is possible to estimate in settings where, in addition to labeled data (i.e., $(X,Y)$ pairs) sampled from the data distribution $\Ptrain \times P_{Y\mid X}$,
we also have access to unlabeled (i.e., $X$ only) data from the target population $\Ptar$. We may  use these two data sets
to train $\wprior$. Although there is no guarantee that the estimate $\wprior$ is accurate, the shape or relative magnitude of $\wprior$ may provide us with useful side information: large values of $\wprior$ can identify portions of the target population
that are \emph{underrepresented} under the data distribution $\Ptrain$.
This motivates us to recalibrate $\wprior$ within the set of density ratios that are isotonic in $\wprior$.

To express this scenario in the notation of the problem formulation above, we assume that the target distribution $\Ptar$ satisfies
an isotonicity constraint, $\Ptar\in\calQ_{\rm iso}(w_0)$, where
\[\calQ_{\rm iso}(w_0) = \left\{ Q: \frac{\mathsf{d} Q}{\mathsf{d}\Ptrain}(x)  \textnormal{ is a monotonically nondecreasing function of }\wprior(x)\right\}.\]
If we assume as before that the target distribution $\Ptar$ satisfies $D_{\rm KL}(\Ptar\| \Ptrain)\leq \rho$, then we can bound
\begin{equation}\label{eq:motivation-delta-bound}\EE_{\Ptar}[R(X)] \leq \EE_{\Ptrain}[R(X)] + \Delta(R; \calQ_{\rm KL}(\rho)\cap \calQ_{\rm iso}(w_0) ).\end{equation}

\paragraph{The benefits of iso-DRL.}
What are the benefits of iso-DRL, as compared to the existing DRL framework?
Of course, thus far the idea is quite straightforward---if we have stronger constraints on $\Ptar$, then 
we can place a tighter bound on the excess risk $\EE_{\Ptar}[R(X)] - \EE_{\Ptrain}[R(X)]$.
But as we will see below, adding the isotonic constraint plays a crucial role in enabling DRL to provide bounds that are useful in practical scenarios.
Specifically, consider a practical setting where the bound $\rho$ on the distribution shift is a positive constant. As we will see
below, the existing worst-case excess risk  $\Delta(R; \calQ_{\rm KL}(\rho))$ of DRL is often quite large, leading to extremely conservative statistical conclusions;
in contrast, the worst-case excess risk $\Delta(R; \calQ_{\rm KL}(\rho)\cap \calQ_{\rm iso}(w_0) )$ given by iso-DRL 
is often vanishingly small, leading to much more informative conclusions.
Moreover, surprisingly, this improvement in the bound does not incur any additional computational challenges---even though the
constraint set $\calQ_{\rm KL}(\rho)\cap \calQ_{\rm iso}(w_0) $ appears more complex than the original set $\calQ_{\rm KL}(\rho)$, we
 will see that $\Delta(R; \calQ_{\rm KL}(\rho)\cap \calQ_{\rm iso}(w_0) )$
can be computed nearly \emph{as easily as} the original quantity $\Delta(R;\calQ_{\rm KL}(\rho))$.
In addition, we further show in Appendix Section~\ref{sec:iso-role} that the worst-case excess risk of iso-DRL can be consistently estimated with noisy observations of $R(X)$, while the estimation of the worst-case excess risk of DRL can be challenging even with bounded risks.

\paragraph{Empirical example: predictive inference for the \texttt{wine quality} dataset.}

\begin{figure}[ht]
    \centering
    \includegraphics[width=0.75\textwidth]{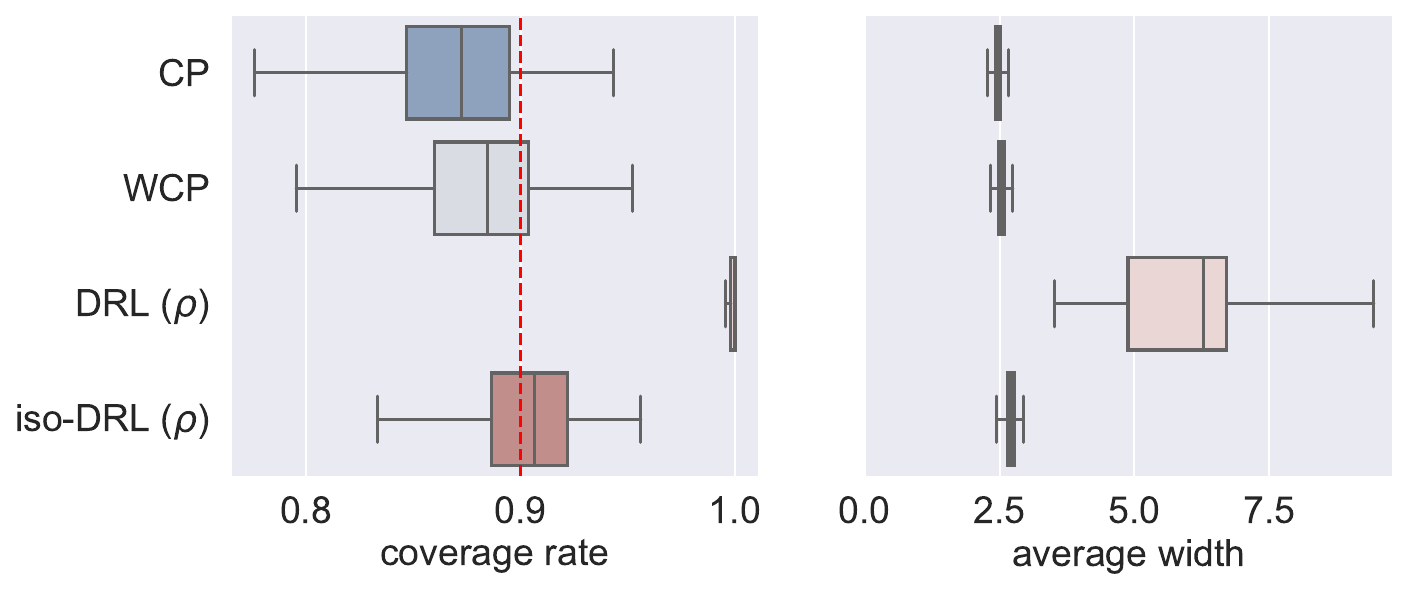}
    \caption{Coverage rate and average width of intervals for the \texttt{wine quality} dataset.
    The red dashed line (in the left-hand plot) marks the nominal coverage level, $1-\alpha = 90\%$. }
    \label{fig:eg1-2}
\end{figure}

To illustrate the advantage of the proposed approach, Figure~\ref{fig:eg1-2} presents a numerical example for a predictive inference
problem on the \texttt{wine quality} dataset \citep{cortez2009modeling}.\footnote{Available at \url{https://archive.ics.uci.edu/dataset/186/wine+quality}} (See Section~\ref{sec:real-simu} for full details of this experiment.)

We are given a pretrained family of prediction bands $\hat C_{1-a}$, indexed by the target coverage level $1-a$. 
At each value $a\in[0,1]$, we define $R_a(X) = \PP(Y\not\in\hat C_{1-a}(X) \mid X)$, the probability of the prediction band failing to cover 
the true response value $Y$ given features $X$. Our goal is to return a prediction band with 90\% coverage---that is, we would like
to choose a value of $a$ such that  the expected risk $\EE_{\Ptar}[R_a(X)] = \PP_{\tPtar}(Y\not\in\hat C_{1-a}(X))$
is bounded by $0.1 =  1 -90\%$. In our experiment, the available data is given by all samples that are white wines (with distribution $\tPtrain$),
while the target population is comprised of the samples that are red wines (with a different distribution $\tPtar$).
In Figure~\ref{fig:eg1-2}, we compare four methods (see Section~\ref{sec:real-simu} for details):
\begin{itemize}
\item An uncorrected interval---using conformal prediction (CP) \citep{vovk2005algorithmic}: the value $a$ is chosen by tuning on the calibration data set (i.e., we choose $a$
to satisfy $\EE_{\Ptrain}[R_a(X)] \leq 0.1$), without correcting
for the distribution shift. 
\item A corrected interval---using weighted conformal prediction (WCP) \citep{Tibshirani2019ConformalPU}: the value $a$ is chosen by tuning on the calibration data set
using an estimated density ratio $\wprior$ to correct for the covariate shift between distributions $\tPtrain$ and $\tPtar$. Since $\wprior$ is estimated from data,
this correction is imperfect.
\item The DRL interval: we choose $a$ to satisfy $\EE_{\Ptrain}[R_a(X)] + \Delta(R_a; \calQ_{\rm KL}(\rho)) \leq 0.1$,
where $\EE_{\Ptrain}[R_a(X)] $ and $\Delta(R_a; \calQ_{\rm KL}(\rho))$ are estimated using the calibration data.
\item The iso-DRL interval: we choose $a$ to satisfy $\EE_{\Ptrain}[R_a(X)] + \Delta(R_a; \calQ_{\rm KL}(\rho)\cap \calQ_{\rm iso}(w_0)) \leq 0.1$,
where $\EE_{\Ptrain}[R_a(X)] $ and $ \Delta(R_a; \calQ_{\rm KL}(\rho)\cap \calQ_{\rm iso}(w_0))$ are estimated using the calibration data.\footnote{For both the DRL and iso-DRL methods, the parameter $\rho$ is an estimate of the actual KL distance $D_{\rm KL}(\Ptar\|\Ptrain)$---see Section~\ref{sec:real-simu} and Appendix~\ref{sec:add-simu_wine} for details}
\end{itemize}

As we can see in Figure~\ref{fig:eg1-2}, the CP and WCP intervals both undercover---for CP, this is because the method does not correct
for distribution shift, while for WCP, this is because the ratio $\wprior$ that corrects for distribution shift is imperfectly estimated.
At the other extreme,
 DRL shows substantial overcoverage with extremely wide prediction intervals due to the worst-case nature of the bound $\Delta(R_a; \calQ_{\rm KL}(\rho))$. In contrast,
our proposed method, iso-DRL, achieves the target coverage rate 90\% without excessive increase in the size of the prediction interval, 
showing the benefit of adding the isotonic constraint to the DRL framework.

The motivating example demonstrates that, when we have access to meaningful---but imperfect---side information (e.g., in the form of the density ratio $\wprior$),
adding the isotonic constraint to iso-DRL can provide an estimate of the risk
that is \emph{more reliable} than a non-distributionally-robust approach, but \emph{less conservative} than the original DRL approach.

\subsection{Organization of paper}\label{sec:org_and_notation}
Section~\ref{sec:setup} introduces a general class of uncertainty sets for candidate distributions and further studies the property of the worst-case excess risk defined in \eqref{eq:opt-0} for generic DRL. For the worst-case excess risk with the isotonic constraint, we prove that it is equivalent to the worst-case excess risk for a projected risk function without the isotonic constraint in Section~\ref{sec:iso}.
In Section~\ref{sec:conv}, we propose an estimator of the worst-case excess risk with the isotonic constraint and establish the estimation error bounds. Numerical results for both synthetic and real data are shown in Section~\ref{sec:simu} and additional related work is summarized in Section~\ref{sec:related}. 
We defer technical proofs and additional simulations to the Appendix.

\paragraph{Notation.}
Before proceeding, we introduce useful notation for theoretical developments later on. 
To begin with, we write $(a)_+$ as the positive part of $a \in \RR$.
We denote by $L_p(\Ptrain)$ ($1 \leq p \leq \infty$) the $L_p$ function space under the probability measure $\Ptrain$, i.e., when $p \neq \infty$,
$L_p(\Ptrain) = \{f:\;\|f\|_p = (\int_{\calX} |f(x)|^p \mathsf{d}P(x))^{1/p} < \infty\}$.
When $p=\infty$, the set $L_{\infty}(P)$ consists of measurable functions that are bounded almost surely under~$\Ptrain$.
In addition, for a measurable function $w$ defined on $\calX$ and a measure $P$ on $\calX$, the pushforward measure $w_{\myhash} P$ denotes the measure satisfiying that $(w_{\myhash} P)(B) = P(w^{-1}(B))$ for any measurable set $B$, where $w^{-1}(B) = \{x \in \calX: w(x) \in B\}$ denotes the preimage of $B$ under $w$. In other words, if $X\sim P$, then $w(X)$ follows the distribution $w_{\myhash}P$. We say a function $h$ is $A_h$-bounded if $\sup_x |h(x)| \leq A_h$.
Fix a partial (pre)order $\orderless$ on $\calX \subseteq \RR^d$. 
A function $g$ is isotonic if $g(x_1) \leq g(x_2)$ for any $x_1 \orderless x_2$.  Correspondingly, we define the cone of isotonic functions by $\Ciso = \{w:  w \text{~is~isotonic}\text{~w.r.t.~partial~order~}\orderless\}$.
Lastly, to compare two probability distributions $Q$ and $P$, the convex ordering $\overset{cvx}{\orderless}$ is defined as $Q^{\prime} \overset{cvx}{\orderless} Q$ if and only if $\EE_{Q^{\prime}} [\psi(X)] \leq \EE_{Q} [\psi(X)]$ for all convex functions $\psi$.

\section{The distributional robustness framework}\label{sec:setup}

As we have explained in Section~\ref{sec:prob}, both the unsupervised and supervised setting under covariate shift can be unified.
Therefore, throughout this section,
 to develop our theoretical results we will use the notation of the unsupervised setting with the risk function $R(X)$, with the understanding that this also covers the supervised setting under covariate shift.

Recall that $\calX$ is the feature domain. 
We consider a bounded risk function $R: \calX \rightarrow [0,B_R]$ with $0 < B_R < \infty$. The goal is to evaluate (or bound) the target risk $\mathbb{E}_{\Ptar} [ R(X)]$ using samples from~$\Ptrain$, by assuming that the target distribution $\Ptar$ is in some sense similar to the available distribution $\Ptrain$---more concretely, by assuming that the target distribution $\Ptar$ lies in some neighborhood $\calQ$ around the distribution $\Ptrain$ of the available data. 

\paragraph{Reformulating the neighborhood.}
To unify the different examples of constraints described in Section~\ref{sec:intro}, we will start by considering
settings where we can  express the constraint
 $Q\in \calQ$ using conditions on the density ratio $w = \mathsf{d} Q/\mathsf{d}\Ptrain$.
This type of framework includes the sensitivity analysis setting via  bounds on $w$ \citep{cornfield1959smoking, rosenbaum1987sensitivity, tan2006distributional, ding2016sensitivity, zhao2019sensitivity,yadlowsky2018bounds, jin2022sensitivity}, and $f$-divergence constraints such as a bound on the KL divergence \citep{duchi2021statistics, namkoong2017variance, duchi2018learning, cauchois2020robust}.

Concretely, we can reparameterize the distribution $Q$ using the density ratio $w(x) = (\mathsf{d} Q/\mathsf{d}\Ptrain)(x)$. 
Then we can reformulate the constraint  $Q \in \calQ$ into a constraint on this density ratio, i.e., $Q\in\calQ \ \Longleftrightarrow \ w_{\myhash} \Ptrain \in\calB$,
where $\calB$ is a set of distributions, and where $w_{\myhash} \Ptrain$ denotes the pushforward measure (as defined in Section~\ref{sec:org_and_notation}).
Let us now consider the two examples mentioned above.

\paragraph{Example 1: bound-constrained distribution shift.}\label{sec:sa_bnd}
In sensitivity analysis, it is common to assume that
the likelihood ratio $ \mathsf{d}\Ptar/\mathsf{d}\Ptrain$ is bounded from above and below.
This corresponds to a constraint set of the form $\calQ = \left\{ Q : a\leq (\mathsf{d}Q/\mathsf{d}\Ptrain)(X)\leq b \textnormal{ $\Ptrain$-almost surely}\right\}$,
for some constants $0 \leq a < 1 < b < +\infty$. In particular, when 
 $a = \Gamma^{-1}$ and $b = \Gamma$ for some $\Gamma > 1$, this constraint set represents the marginal $\Gamma$-selection model for the density ratio in sensitivity analysis \citep{rosenbaum1987sensitivity,tan2006distributional}. 
By defining
\[
\calB = \calB_{a,b} = \left\{\tilde Q:\;\EE_{Z\sim\tilde Q}[Z]=1,\; \PP_{Z\sim \tilde Q}(a \leq Z \leq b)=1\right\},
\]
we can verify that
$Q\in\calQ \ \Longleftrightarrow \ w_{\myhash} \Ptrain \in\calB_{a,b}$ with $w(x) =  (\mathsf{d} Q/\mathsf{d}\Ptrain)(x)$.

\paragraph{Example 2: \texorpdfstring{$f$}{f}-divergence constrained distribution shift.}\label{sec:f-div}
The $f$-divergence is a generalized way of measuring the distance between distributions, which includes common  metrics such as KL divergence or chi-squared divergence as special cases. 
For a convex function $f:[0,\infty)\to\mathbb{R}$ satisfying $f(1) = 0$, the $f$-divergence \citep{ali1966general,renyi1961measures} of $Q$ from $\Ptrain$ is defined as
$D_f\left(Q || \Ptrain\right) = \EE_{\Ptrain}[f((\mathsf{d}Q/\mathsf{d}\Ptrain)(X))]$.
In this example, we consider a constraint set $\calQ$ defined via a bound on the $f$-divergence: $\calQ = \left\{ Q : D_f\left(Q || \Ptrain\right)\leq \rho\right\}$.
For instance, if we take $\calQ = \calQ_{\rm KL}(\rho) = \{Q : D_{\rm KL}(Q\| P)\leq \rho\}$,
{this corresponds to choosing $f(x) = x\log(x)$.}
Choosing 
\[
\calB = \calB_{f,\rho} = \{\tilde Q:\; \EE_{Z\sim\tilde Q}[Z]=1,\; \EE_{Z\sim \tilde Q}[f(Z)] \leq \rho,\;\PP_{Z\sim \tilde{Q}}(Z \geq 0)=1\},
\]
we can verify that $Q\in\calQ \ \Longleftrightarrow \ w_{\myhash} \Ptrain \in\calB_{f,\rho}$ with $w(x) =  (\mathsf{d} Q/\mathsf{d}\Ptrain)(x)$.

\subsection{Worst-case excess risk with DRL}\label{sec:no-cone}
In this section,
we explore some properties of the generic DRL, without the isotonic constraint. Building this framework will help us to introduce the isotonic constraint as follows. 

Based on the equivalence of $\calQ$ and $\calB$ in representing the uncertainty set, we focus on the following equivalent representation of $ \Delta(R;\calQ)$: 
\begin{align}\label{opt:unconst}
\Du \quad = \quad \sup_{w \geq 0} \qquad & \EE_{\Ptrain}\left[w(X)R(X)\right] - \alphaP  \nonumber\\*
{\rm subject~to} \qquad &w_{\myhash} \Ptrain\in \calB,
\end{align}
 where abusing notation we now write $\Delta(\cdot ;\calB)$ to express that $\calB$ is a constraint on the distribution of the density ratio $w(X)=(\mathsf{d}Q/\mathsf{d}\Ptrain)(X)$,
where previously we instead wrote $\Delta(\cdot;\calQ)$.
We will say that $\Du$ is \emph{attainable} if this supremum is attained by some $w^*$ in the constraint set.
Throughout the paper, we assume that the set of distributions $\calB$ satisfies the following condition. 
\begin{condition}\label{cond:B-cond}
{The set $\calB$ contains the point mass on the value $1$. Moreover, $\calB$} is closed under convex ordering, that is, if $Q \in \calB$, then for any $Q^{\prime} \overset{cvx}{\orderless} Q$, it holds that $Q^{\prime} \in \calB$.
\end{condition}
This condition enables the following reformulation of the quantity of interest, $\Delta(R;\calB)$:
\begin{proposition}\label{prop:ucons-kkt}
Assume Condition~\ref{cond:B-cond} holds. Then $\Du$ can be written as
\begin{align*}
\Du \quad = \quad \sup_{\phi: \RR \rightarrow \RR_+} \qquad & \EE_{\Ptrain}\left[(\phi \circ R)(X)R(X)\right] - \alphaP  \nonumber\\
{\rm subject~to} \qquad &(\phi \circ R)_{\myhash} \Ptrain\in \calB, \qquad \phi\text{~is~nondecreasing}.
\end{align*}
\end{proposition}
Moreover, if $\Du$ is attainable (i.e., the supremum is attained by some $w^*$ satisfying the constraints), then
this equivalent formulation is attainable as well (i.e., the supremum is attained by some $\phi^*$ satisfying the constraints), and it then holds that $w^*(X) = \phi^*(R(X))$ $\Ptrain$-almost surely.

In words, this proposition shows that the excess risk is maximized by 
considering functions $w(x)$ that are monotonically nondecreasing with respect to $R(x)$.
This is intuitive, since maximizing the expected value of $w(X)R(X)$ implies that we should choose a function $w$ that is large when $R$ is large.
Most importantly, Proposition~\ref{prop:ucons-kkt} implies that for a class of constraint sets $\calB$, the optimal value in the constrained optimization problem~\eqref{opt:unconst} 
only depends on the distribution of $X$ through the distribution of $R(X)$. 
We note that, in the special case when $\calB$ is specified in terms of an $f$-divergence (as in Example 2 above), the conclusion of Proposition~\ref{prop:ucons-kkt} is established by \citealp{donsker1976asymptotic, lam2016robust, namkoong2022minimax}.
We verify that this result holds in aforementioned settings as follows.

\paragraph{Returning to Example 1: bound-constrained distribution shift.}\label{sec:sa_bnd_}
Recall that in this example, we take the constraint set
$
\calB = \calB_{a,b} = \{\tilde Q:\;\EE_{Z\sim\tilde Q}[Z]=1,\; \PP_{Z\sim \tilde Q}(a \leq Z \leq b)=1\},
$
for some $0 \leq a < 1 < b < +\infty$.
It is straightforward to verify that $\calB_{a,b}$ satisfies Condition \ref{cond:B-cond}, implying that Proposition~\ref{prop:ucons-kkt}
can be applied.

Moreover, we can actually calculate
the maximizing density ratio $w^*(x)$ explicitly.
The worst-case density ratio that attains the worst-case excess risk takes the form
\[
w^*(x) = a \cdot \ind\left\{R(x) < q_R\left(\tfrac{b-1}{b-a}\right)\right\} + b \cdot \ind\left\{R(x) >q_R\left(\tfrac{b-1}{b-a}\right) \right\} + c\cdot \ind\left\{R(x) = q_R\left(\tfrac{b-1}{b-a}\right)\right\},
\]
where $q_R(t)$ is the $t$-quantile of the distribution of $R(X)$ under $X\sim \Ptrain$ and $c\in[a,b]$ is defined as the unique value ensuring that $\EE[w^*(X)]=1$, namely,
\[
c = a + \frac{(b-a) t^* - (b-1)}{\PP\left\{R(X) = q_R\left(\frac{b-1}{b-a}\right)\right\}} \qquad \text{with} \qquad t^* = 
\PP\left\{R(X) \leq  q_R\left(\tfrac{b-1}{b-a}\right)\right\} \geq \frac{b-1}{b-a}.
\]
We can see that $w^*(x)$ is nondecreasing in $R(x)$, 
validating the conclusion of Proposition~\ref{prop:ucons-kkt}.

\paragraph{Returning to Example 2: \texorpdfstring{$f$}{f}-divergence constrained distribution shift.}\label{sec:f-div-}
Recall that for an $f$-divergence constraint, we define
$
\calB = \calB_{f,\rho} = \{\tilde Q:\; \EE_{Z\sim\tilde Q}[Z]=1,\; \EE_{Z\sim \tilde Q}[f(Z)] \leq \rho,\;Z \geq 0\}.
$
Since $f$ is convex, this immediately implies that $\calB_{f,\rho}$ satisfies Condition \ref{cond:B-cond}. 
If we further assume that $f$ is differentiable, by the results of \cite{shapiro2017distributionally,donsker1976asymptotic, lam2016robust}, the worst-case excess risk $\Df$ is attained at
\[
w^*(x) = w(x; \lambda^*,\nu^*) = \left\{(f^\prime)^{-1}\left(\frac{R(x) - \nu^*}{\lambda^*}\right)\right\}_+,
\]
where $\lambda^*$, $\nu^*$ are the solutions to the dual problem
\begin{align}
\inf_{\lambda \geq 0, \;\nu} \bigg\{\lambda \rho + \nu + \EE_{\Ptrain}\bigg[w(X;\lambda,\nu)(R(X) - \nu) - \lambda f(w(X;\lambda,\nu))\bigg]\bigg\}.
\end{align}
Since $f$ is convex, its inverse derivative $(f^\prime)^{-1}$ is nondecreasing, 
meaning that $w^*(x)$ is nondecreasing in $R(x)$, which again validates the result in Proposition~\ref{prop:ucons-kkt}.

\section{Worst-case excess risk with an isotonic constraint}\label{sec:iso}
In this section, we will now formally introduce our iso-DRL method, adding an isotonic constraint
to the DRL framework developed in Section~\ref{sec:setup} above. As in Section~\ref{sec:setup}, throughout this section we use the notation
of the unsupervised learning setting with risk $R(X)$, since the supervised case can also be reduced to this setting.

Recall the cone of isotonic functions
$\Ciso = \{w:\calX\to \RR \,:\; w \text{~is~isotonic}\text{~w.r.t.~}\orderless\}$.
In this paper, we actually allow $\orderless$ to be a partial \emph{preorder} rather than a partial order, meaning
that it may be the case that both $x\orderless x'$ and $x'\orderless x$, even when $x\neq x'$. As an example, we denote $\Cisow = \{w: w(x) \textnormal{ is a monotonically nondecreasing function of }\wprior(x)\}$---this is obtained by the (pre)order given by $x\orderless x'$ whenever $w_0(x)\leq w_0(x')$. 

Our focus is the worst-case excess risk with the isotonic constraint: 
\begin{align}\label{eq:opt-iso0}
\Diso \quad = \quad \sup_{w \geq 0} \qquad & \EE_{\Ptrain}\left[w(X)R(X)\right] - \alphaP  \nonumber\\*
{\rm subject~to} \qquad &  w_{\myhash} \Ptrain \in \calB,\quad w\in \Ciso.
\end{align} 
To make this more concrete with a specific example, in the bound~\eqref{eq:motivation-delta-bound}, this example corresponds to choosing $\calB = \calB_{f,\rho}$ for the $f$-divergence $f(x) = x\log x$, as for the KL distance constraint. In particular, the bound~\eqref{eq:motivation-delta-bound}
assumed two constraints on the distribution $\Ptar$---first, $D_{\rm KL}(\Ptar\|\Ptrain)\leq\rho$ (which corresponds
to assuming $(\mathsf{d}\Ptar/\mathsf{d}\Ptrain)_{\myhash}\Ptrain\in\calB_{f,\rho}$, in our new notation), and second,  $\Ptar \in\calQ_{\rm iso}(w_0)$
(which is expressed by assuming $w\in\Ciso$ when we take the partial (pre)order
defined as $x\orderless x'$ whenever $\wprior(x)\leq \wprior(x')$---or equivalently, we can write this as $w\in\Cisow$).

\subsection{Equivalent formulation}\label{sec:equiv}

Optimization problems with isotonic constraints may be difficult to tackle both theoretically and computationally,
since the isotonic cone, despite
being convex,  may be challenging to optimize over when working with an infinite-dimensional object such as the density ratio.
 In this section, we will show that the maximization problem~\eqref{eq:opt-iso0}
 can equivalently be reformulated as an optimization problem \emph{without} an isotonic constraint, by drawing a connection to
 the original (not isotonic) DRL maximization problem~\eqref{opt:unconst}.

Given the probability measure $\Ptrain$, we will define $\pi$ as the projection to the isotonic cone $\Ciso$ with respect to $L_2(\Ptrain)$, i.e., $\pi(a) = \argmin {b \in \Ciso} \int (a(x) - b(x))^2 \mathsf{d}\Ptrain(x)$.
As $L_2(\Ptrain)$ is reflexive and strictly convex, the projection $\pi(a)$ exists and is unique (up to sets of measure zero) for all $a \in L_2(\Ptrain)$ \citep{megginson2012introduction}. Then, with the projection $\pi$ in place, we
are ready to state our main equivalence result.
\begin{theorem}\label{thm:equiv}
For any $\calB$ and any partial (pre)order $\orderless$ on $\calX$, it holds that $\Diso \leq \DRiso$.
If in addition Condition~\ref{cond:B-cond} holds, then we have \[\Diso = \DRiso,\]
and moreover, $\Diso$ is attainable if and only if $\DRiso$ is attainable.
\end{theorem}


To interpret this theorem, recall from the definition~\eqref{opt:unconst} that we have
\begin{align}\label{opt:proj}
\DRiso \quad = \quad \sup_{w \geq 0} \quad & \EE_{\Ptrain}\big[w(X) [\pi(R)](X)\big] - \EE_{\Ptrain}\big[[\pi(R)](X)\big]  \nonumber\\*
{\rm subject~to} \quad & w_{\myhash} \Ptrain \in \calB.
\end{align}
Compared with the formulation~\eqref{eq:opt-iso0} that defines the isotonic worst-case risk $\Diso$,
we see that this equivalent formulation removes the constraint $w\in\Ciso$ by replacing $R$ with its isotonic
projection $\pi(R)$. 
This brings computational benefits.  The equivalent formulation~\eqref{opt:proj} separates two constraints $w_{\myhash}\Ptrain \in \calB$ and $w \in \Ciso$, 
allowing us to first project the risk function $R$ onto $\Ciso$, and then solve a problem that is as simple as the problem stated earlier in~\eqref{opt:unconst}.
More concretely, 
as seen 
in Examples 1 and 2, for many common choices of $\calB$, we have closed-form solutions to~\eqref{opt:proj} in terms of the projected risk $\pi(R)$.

\subsection{Setting: iso-DRL with estimated density ratio}\label{sec:recalib}
We now return to the scenario described in~\eqref{eq:motivation-delta-bound} in Section~\ref{sec:intro-iso},
where we would like to recalibrate a pretrained density ratio $\wprior$ that estimates $\mathsf{d}\Ptar/\mathsf{d}\Ptrain$.
As the shape or relative magnitude of $\wprior$ could contain useful information about the true density ratio, we consider candidate distributions with the density ratio as an isotonic function of $\wprior$, which is equivalent to considering the partial (pre)order $x\orderless x' \ \Longleftrightarrow \ \wprior(x)\leq \wprior(x')$. We will denote the specific isotonic cone under this partial (pre)order as $\Cisow$ and its isotonic projection as $\pi_{\wprior}$, and abusing notation, we write
$\Disow$ to denote the excess risk for this particular setting, to emphasize the role of $\wprior$. Under the condition of Theorem~\ref{thm:equiv}, we have the equivalence $\Disow= \Delta(\pi_{w_0}(R);\calB)$.
To understand the projection onto the cone $\Cisow$ in a more straightforward way, we can derive a further simplification.
Write $\pi_1$ to denote the isotonic projection of functions $\RR\to\RR$ under the measure $(\wprior)_{\myhash}\Ptrain$,
and define a function $\tilde R:\RR\to\RR$ to satisfy $\tilde R(\wprior(X)) = \EE_{\Ptrain}[R(X)\mid \wprior(X)]$
$\Ptrain$-almost surely. 
We then have the following simplified equivalence:
\begin{proposition}\label{prop:equi-rewt}
Assume Condition~\ref{cond:B-cond} holds. We have the equivalence $\Disow = \altDwiso$,
where we define
\begin{align}\label{eq:wt-recalib}
\Dw \quad = \quad \sup_{h:~h \circ \wprior \geq 0} \qquad & \EE_{\Ptrain}\left[(h \circ \wprior)(X)R(X)\right] - \alphaP  \nonumber\\*
{\rm subject~to} \qquad & \left(h \circ \wprior\right)_{\myhash} \Ptrain \in \calB.
\end{align} 
\end{proposition}
In comparison to the equivalence $\Disow= \Delta(\pi_{w_0}(R);\calB)$, 
the equivalence in the proposition relies on an isotonic projection 
with respect to the canonical order on the real line (i.e., the projection $\pi_1$). 

Moreover, when the true distribution shift does not obey the isotonic constraint exactly, in Appendix Section~\ref{sec:mis-iso}, we can nonetheless provide a bound on the worst-case excess risk, which is tighter than the (non-iso) DRL bound whenever the isotonic constraint provides a reasonable approximation.

\section{Estimation of worst-case excess risk with isotonic constraint}\label{sec:conv}
So far, our focus has been on the population level problem, namely, we have assumed full access to the data distribution $\Ptrain$ and the risk function $R$. 
In practice, however, we may only be able to access the data distribution $\Ptrain$ via samples, and we may only be able to learn about the risk function~$R$ via noisy evaluations of $R(X)$ on each sampled point $X$ in the unsupervised setting. Or, in the supervised setting, we can only access $\tPtrain$ via
samples of labeled data points drawn from this distribution, and can learn about~$r$ only through evaluating $r(X,Y)$ on these sampled data points.

In this section, we propose a fully data dependent estimator for the worst-case excess risk $\Diso$. 
Moreover, we characterize the estimation error for different choices of $\calB$, including the bounds constraint and the $f$-divergence constraint for the distribution shift.

\subsection{Plug-in estimators}
We start with presenting the plug-in estimators of the worst-case excess risk under the isotonic constraint in both the unsupervised and supervised settings.  
\paragraph{The unsupervised setting.} We have $n$ i.i.d.~observations $\{X_i\}_{i\leq n}$ from a  distribution $P$. Given a risk function $R : \mathcal{X} \to \mathbb{R}_{+}$, and the uncertainty set  $\mathcal{B}$, 
we estimate the worst-case excess risk $\Diso$ (cf.~Equation~\eqref{eq:opt-iso0}) via
 \begin{align}\label{eq:opt-est-no-noisy}
\hatDiso \quad\coloneqq\quad \max_{w \geq 0} \qquad & \frac{1}{n} \sum_{i \leq n} w(X_i) R(X_i) - \frac{1}{n}\sum_{i \leq n} R(X_i)  \nonumber\\*
{\rm subject~to} \qquad & w_{\myhash} \hat P_n \in \calB, \quad w \in \Ciso.
\end{align}
Here, $\hat P_n$ denotes the empirical distribution of the sample $\{X_i\}_{i \leq n}$ drawn from $P$.

\paragraph{The supervised setting.}
In this case, we have $\{(X_i,Y_i)\}_{i \leq n}$ drawn i.i.d.~from $\tPtrain = \Ptrain \times P_{Y \mid X}$. Given a risk function $r$, and the uncertainty set $
\calB$, we propose to estimate the worst-case excess risk $\hatDisor$ by replacing $R(X_i)$ with $r(X_i,Y_i)$ in \eqref{eq:opt-est-no-noisy}.

\subsubsection{Adding a boundedness constraint}
When calculating the excess risk at the  population level, the constraint set $\calB$ may not require $w$ to be bounded---specifically,
while $\calB_{a,b}$ imposes an upper bound on $w$, the $f$-divergence constraint set $\calB_{f,\rho}$ does not. 
In the empirical setting, however, a boundedness constraint is more crucial: we want to avoid degenerate scenarios,
such as $w(X_i)$ taking an arbitrarily large value for a single $i$, and being zero for the remaining $n-1$ data points.
To this end, we will assume from this point on that $\calB$ includes a boundedness constraint:
\begin{condition}\label{cond:bnd-w}
There exists $\wbnd$ such that any distribution $Q\in\calB$ is supported on $[0,\wbnd]$.
\end{condition}
This is trivially true for $\calB=\calB_{a,b}$ with $\wbnd = b$. But this constraint actually allows us to work
 with the $f$-divergence example, as well, as established by the following result. 
 \begin{proposition}\label{thm:thirdterm_2}
Assume the convex function $f$ is differentiable on $\RR_+$.
The worst-case excess risk $\Dfiso$ is attained at some $\wfiso \in \Ciso$ with $\|\wfiso\|_{\infty} < \infty$.
\end{proposition}
In particular, defining $\calB_{f,\rho,\wbnd} = \{\tilde Q:\; \EE_{Z\sim\tilde Q}[Z]=1,\; \EE_{Z\sim \tilde Q}[f(Z)] \leq \rho,\;\PP_{Z \sim \tilde Q}(0 \leq Z \leq \wbnd) = 1\}$,
which adds a boundedness requirement in addition to the $f$-divergence constraint, we can see that for sufficiently large $\wbnd$ (namely, $\wbnd \geq \|\wfiso\|_\infty$), even though $\calB_{f,\rho,\wbnd} \subsetneq \calB_{f,\rho}$, it nonetheless holds that $\DfisoG = \Dfiso$.
Therefore, by working with the constraint set $\calB_{f,\rho,\wbnd}$, we are estimating the \emph{same} excess risk,
but Condition~\ref{cond:bnd-w} nonetheless holds. (Of course, in practice, the value of $\|\wfiso\|_{\infty}$ is unknown
and so we can simply set $\Omega$ to be a large constant.)

\subsection{Computation: estimation after projection}\label{sec:est_proj}
Before moving onto the statistical performance of the two estimators $\hatDiso$ and $\hatDisor$, we pause to discuss fast computational methods for these. 
The key is Theorem~\ref{thm:equiv}---we may accelerate the computation of both estimators via an equivalent optimization problem without the isotonic constraint. 
  
To be more specific, we begin by considering the supervised setting. Denote $r^{\iso} = (r^{\iso}_i)_{i \leq n} \in \RR^n$ as the isotonic projection of $(r(X_i,Y_i))_{i \leq n}$ with respect to the empirical distribution $\hat P_n$ under the partial order $\orderless$. Then, consider the optimization problem
\begin{align}\label{eq:opt-est-noisy-proj}
\hatDriso \quad\coloneqq\quad \max_{w \geq 0} \qquad & \frac{1}{n} \sum_{i \leq n} w(X_i) r^{\iso}_i - \frac{1}{n}\sum_{i \leq n} r^{\iso}_i  \nonumber\\*
{\rm subject~to} \qquad & w_{\myhash} \hat P_n \in \calB.
\end{align}
By Theorem~\ref{thm:equiv} (applied with $\hat P_n$ in place of $\Ptrain$), we have $\hatDriso = \hatDisor$. 
Analogously, in the unsupervised setting, we instead have $\hatDiso = \hatDRiso$,
where $R^{\iso} = (R^{\iso}_i)_{i \leq n} \in \RR^n$ as the isotonic projection of $(R(X_i))_{i \leq n}$ with respect to the empirical distribution $\hat P_n$ under the partial order $\orderless$.

Note that in iso-DRL with estimated density ratio in Section~\ref{sec:recalib}, we can simply apply the isotonic regression for $(r(X_i,Y_i))_{i \leq n}$ on $(\wprior(X_i))_{i \leq n}$ to obtain the projected risk. 
We can now see concretely that this equivalence allows for a much more efficient calculation.
For example, in the case $\calX=\mathbb{R}$, this isotonic projection can be computed in $\mathcal{O}(n)$ time (e.g., via the PAVA algorithm, which 
provides an exact calculation of isotonic projection in $\mathbb{R}^n$ \citep{grotzinger1984projections}), this leads to a very simple implementation for computing 
$\hatDriso$---in particular, once the vector $r^{\iso}$ has been computed, the remaining optimization problem
is simple since there is no remaining isotonic constraint.

\subsection{Performance guarantees for plug-in estimators}\label{sec:empirical_theory}
In this section, we present the performance guarantees for plug-in estimators for a general constraint set $\calB$.
To jump ahead to the conclusion, we will see that our results imply the following consistency properties for the settings $\calB=\calB_{a,b}$
and $\calB=\calB_{f,\rho,\wbnd}$. 
Here we define the Rademacher complexity of a function class $\calG$ by $\calR_n(\calG) = \EE[\sup_{g \in \calG} |n^{-1} \sum_{i \leq n} \sigma_i g(Z_i)|]$,
where $\{Z_i\}_{i \leq n}$ is a sample of size $n$ from $\Ptrain$ and $\{\sigma_i\}_{i \leq n}$ are independent random variables drawn uniformly
from $\{+1,-1\}$.
\begin{proposition}[Informal result for examples]\label{cor:calB_examples} For both $\calB = \calB_{a,b}$ and $\calB = \calB_{f,\rho,\wbnd}$, and for both supervised ($\hatDisoU = \hatDisor$) and unsupervised ($\hatDisoU = \hatDiso$) learning, and under some mild additional conditions specified below, it holds with probability  $\geq 1- 3n^{-1}$ that
\[ \left|\hatDisoU - \Diso\right|\leq C\left( \calR_n(\CisoB) + \sqrt{\frac{\log n}{n}}\right),\]
where $\CisoB = \{w \in \Ciso:\;0 \leq w \leq \wbnd\}$ is the bounded isotonic cone,  where the constant $C$ will be defined in the theorems below, and where we set $\Omega=b$ for the case $\calB=\calB_{a,b}$.
\end{proposition}
Our bounds rely on the Rademacher complexity term $\calR_n(\CisoB)$, which will naturally
depend on the properties of the (pre)ordering $\orderless$ that defines this isotonic cone. To provide further intuition, we
now give two concrete examples to make apparent the dependence of the Rademacher complexity on the sample size. 
\begin{enumerate}
\item[(1)] When $d=1$, e.g., in the setting of density ratio recalibration in Section~\ref{sec:recalib}, similar to the results of \cite{chatterjee2019adaptive}, one can show by Dudley's theorem \citep{dudley1967sizes} that $\calR_n\left(\CisoB\right) \lesssim n^{-1/2}$ up to logarithmic factors.
\item[(2)] For $\RR^d$ with a fixed dimension $d \geq 2$ and a bounded domain $\calX$ equipped with the componentwise order \citep{han2019isotonic,deng2020isotonic,gao2007entropy}, i.e., $x \orderless z$ if and only if $x_j \leq z_j$ for all $j \in [d]$, by \cite{han2019isotonic}, if the density $\mathsf{d}P(x)$ is bounded below (away from zero) and above, then we have $\calR_n\left(\CisoB\right) \lesssim n^{-1/d}$ up to logarithmic factors.
\end{enumerate}

\subsubsection{Formal results}
Now we turn to developing these results formally, in a general framework. 
We will begin with a deterministic result, which shows that, if certain concentration inequalities hold,
then $\hatDbndisoU$ is an accurate estimate of $\Dbndiso$. Then we will show that the concentration results hold with high probability,
in both of our two settings, $\calB=\calB_{a,b}$
and $\calB=\calB_{f,\rho,\wbnd}$.

We first need a few definitions. For any distributions $P_0,P_1$, if $w_{\myhash}P_0\in\calB$, we define
\[\vareps_{\calB} \left(w ; P_0, P_1\right)  =\inf\left\{ s\geq  0 \ : \  \exists \ t\geq 0, \, \big((1-s)\cdot w + t\cdot \mathbf{1}\big)_{\myhash}P_1\in\calB\right\}.\]
In other words, if weight function $w$ satisfies the constraints relative to distribution $P_0$,
we need to find  constants $s,t$ such that the modified weight function $(1-s)\cdot w + t\cdot \mathbf{1}$ satisfies
the constraints relative to distribution $P_1$. (Note that we must have $\vareps_{\calB} \left(w ; P_0, P_1\right) \leq 1$, since choosing $s=t=1$ will always be feasible, because $\mathbf{1}_{\myhash}P_1$ is the point mass on the value 1, and therefore satisfies the constraints of $\calB$, by assumption.)
Of importance is the quantity
\[\vareps_{\calB} = \sup_{w\in\CisoB} \max\left\{ \vareps_{\calB}\left(w; \Ptrain,\hat P_n\right),  \vareps_{\calB}\left(w; \hat P_n,\Ptrain\right)\right\}, \]
which characterizes the feasibility gap between the population and sample problems.
In addition, we define 
\[\vareps_R = \sup_{w\in\CisoB} \left|\EE_{\hat P_n}[(w(X)-1)r(X,Y)] - \EE_{\Ptrain}[(w(X)-1)R(X)]\right|\]
in the case of unsupervised learning, or
\[\vareps_R = \sup_{w\in\CisoB} \left|\EE_{\hat P_n}[(w(X)-1)R(X)] - \EE_{\Ptrain}[(w(X)-1)R(X)]\right|\]
in the case of supervised learning. 
The value of $\vareps_R$ measures the concentration between the empirical risk and the population one. 
With these definitions in place, we are ready to state the generic performance guarantee of the plug-in estimators. 
\begin{theorem}\label{thm:deterministic}
Suppose that the risk is $B_R$-bounded (i.e., $R$ or $r$, in the unsupervised or supervised case, respectively),
and that the constraint set $\calB$ satisfies  Condition~\ref{cond:bnd-w}.
Then, it holds for both supervised ($\hatDisoU = \hatDisor$) and unsupervised ($\hatDisoU = \hatDiso$) learning that
\[\left|\hatDisoU - \Diso\right|\leq  \vareps_R + 2B_R\wbnd \cdot \vareps_{\calB}.\]
\end{theorem}

Of course, in order for this result to be meaningful, we need to ensure that $\vareps_R$ and $\vareps_{\calB}$ are likely to be small, with high probability,
We now turn to the question of establishing such concentration results.  First we bound $\vareps_R$.
\begin{lemma}\label{lem:vareps_R}
Suppose that the risk is $B_R$-bounded (i.e., $R$ or $r$, in the unsupervised or supervised case, respectively). 
Then, with probability at least $1-n^{-1}$, it holds that $\vareps_R \leq 4B_R\calR_n(\CisoB) + B_R\wbnd \sqrt{\log n/(2n)}$.
\end{lemma}

Next we turn to bounding $\vareps_{\calB}$, which we will do separately for our two examples. 
\begin{lemma}\label{lem:vareps_calB_a_b}
Let $\calB = \calB_{a,b}$, where $a<1<b$.
Then, with probability at least $1-n^{-1}$, it holds that $\vareps_{\calB} \leq  C(\calR_n(\CisoB) + \Omega \sqrt{\log n/(2n)})$,
where we take $\Omega=b$ and $C$ depends only on $a,b$.
\end{lemma}
Finally, to complete this section, we turn to the $f$-divergence constraint, $\calB_{f,\rho}$.
\begin{lemma}\label{lem:vareps_calB_f_rho}
Let $\calB = \calB_{f,\rho,\wbnd}$, where we take any $\wbnd\geq \|\wfiso\|_{\infty}$ for $\wfiso$ defined as in Proposition~\ref{thm:thirdterm_2}.
Assume also that $f$ is $L_\wbnd$-Lipschitz on $[0,\wbnd]$. Then, with probability at least $1-2n^{-1}$, it holds that $\vareps_{\calB} \leq C ( \calR_n(\CisoB) + \sqrt{\log n/(2n)})$,
where $C$ depends only on $\wbnd$, $L_\wbnd$, and $\rho$.
\end{lemma}

\subsection{The role of the isotonic constraint}\label{sec:iso-role-main}
The consistency bounds developed above show that, under appropriate conditions,
the error in estimating $\Diso$ can be controlled whenever the appropriate Rademacher complexity terms are small. In the Appendix Section~\ref{sec:iso-role}, we will construct an example with $\calB=\calB_{a,b}$ such that, without an isotonic constraint, $\Dbnd = 0$ but $\hatDbndr > c$ with high probability, where $c > 0$ doesn’t vanish with $n$,
i.e., this empirical estimate is \emph{not} a consistent estimator of the true excess risk. 
This suggests that the isotonic constraint plays an important role: essentially, the isotonic constraint induces a form of regularization, ensuring that we work with a low-complexity class of functions.

\section{Numerical experiments}\label{sec:simu}
In this section, we demonstrate the benefits of iso-DRL in calibrating prediction sets under covariate shift with empirical examples,
as previewed in Section~\ref{sec:intro-iso}. 
Throughout all experiments, we have a training data set $\calD_{\tr}$ containing data points $(X_i,Y_i)$ drawn from the data distribution $\tPtrain$,  and a test set $\calD_{\te}$ containing data points $(\tilde X_i,\tilde Y_i)$  drawn from $\tPtar$. 
We consider both synthetic and real datasets. Code to reproduce all experiments is available at \url{https://github.com/yugjerry/iso-DRL}.

\paragraph{Background.}
When covariate shift is present, \cite{Tibshirani2019ConformalPU} proposes the weighted conformal prediction (WCP) 
method, which produces a prediction set $C^{\wprior}_{1-\alpha}(X)$ with an estimated density ratio $\wprior$, which is only valid for the shifted covariate distribution $\hat P$ defined by $\mathsf{d}\hat P \propto \wprior \cdot \mathsf{d}\Ptrain$. 
The validity for the target distribution $\tPtar$ is only guaranteed up to a coverage gap due to the estimation error or potential misspecification in $\wprior$ \citep{lei2020conformal,candes2023conformalized,gui2024conformalized,gui2023conformalized}.

\paragraph{Dataset partition.}
The datasets $\calD_{\tr}$ and $\calD_{\te}$ are partitioned as follows: (1) first, we use a subset $\calD_1\subseteq \calD_{\tr}$ of the training data of size $|\calD_1| = n_{\rm pre}$, and a subset $\calD_{\te,1}\subseteq \calD_{\te}$ of the test data of size $|\calD_{\te,1}| = n_{\rm pre}$,
to train the function $\wprior$; (2) next, we use a subset  $\calD_2\subseteq \calD_{\tr}\backslash \calD_1$ of the training data of size $|\calD_2| = n_{\tr}$
to train CP or WCP prediction intervals; (3) then, $\calD_3 = \calD_{\tr}\backslash (\calD_1 \cup \calD_2)$ is used to for estimating upper bounds on the excess risk for the DRL and iso-DRL methods. We further define $n = |\calD_3|$ to ease notations; (4) finally, $\calD_{\te,0} = \calD_{\te}\backslash \calD_{\te,1}$  is used for estimating the actual performance of each method relative to the target distribution: defining $ n_{\te}=|\calD_{\te,0}|$, we compute $\texttt{Coverage~rate}(C, \alpha) = n_{\te}^{-1}\sum_{i \in \calD_{\te,0}} \ind\{\tilde Y_i \in C(\tilde X_i)\}$.
We next turn to the details of how these steps are carried out.

\paragraph{Initial density ratio estimation.}
Using data from $\calD_1$ and $\calD_{\te,1}$, we construct a data set comprised of the covariate $X$ 
and a binary label $L \in \{0,1\}$ ($0$ for the training points, $1$ for the test points).
We then fit a logistic regression model and obtain the estimated probability $\hat p(x)$ for $\PP(L = 1 \mid X = x)$, with which we define
$\wprior(x) = \hat p(x)/(1 - \hat p(x))$.
\paragraph{Split conformal prediction and weighted split conformal prediction.}
With data from $\calD_2$, we use Ordinary Least Squares (OLS) as the base algorithm, where we denote $\hat \mu$ as the fitted regression model, and apply split conformal prediction with the nonconformity score $V(x,y) = |y - \hat \mu(x)|$
to obtain the following prediction intervals for comparison: (1) CP: conformal prediction interval $C_{1-\alpha}$ without adjusting for covariate shift; (2) WCP-oracle: weighted conformal prediction interval $C^{w^*}_{1-\alpha}$ with true density ratio $w^* = \mathsf{d}\Ptar/\mathsf{d}\Ptrain$; (3) WCP: weighted conformal prediction interval $C^{\wprior}_{1-\alpha}$ with estimated density ratio $\wprior$.

\paragraph{DRL methods: estimation of worst-case excess risks.}
We then consider two distributionally robust methods.
Using the subset $\calD_3$ of the training data, the observed risks can be calculated by $r_i = \ind\left\{Y_i \notin C_{1-\alpha}(X_i)\right\}$, $i\in\calD_3$.
We adopt the KL divergence constraint $D_{\rm KL}(Q \| \Ptrain) \leq \rho$ to measure the magnitude of distribution shift. Then, 
we obtain 
\begin{align}\label{eq:delta-hat-DRL}
\hat \Delta(\alpha) = & \max_{\Vert w\Vert_{\infty} \leq \wbnd} \quad \frac{1}{n} \sum_{i \in \calD_3} (w_i-1) r_i \quad {\rm s.t.}\quad \frac{1}{n}\sum_{i \in \calD_3} w_i=1, \;\frac{1}{n}\sum_{i \in \calD_3} w_i \log w_i \leq \rho,
\end{align}
with the upper bound set as $\wbnd=100$ throughout the experiments.
Next, given the estimated density ratio $\wprior$, we run isotonic regression for $(r_i)_{i \leq n}$ on $(\wprior(X^{(3)}_i))_{i \leq n}$ to obtain the projected risk $(r^\iso_i)_{i \in \calD_3}$, with which we can calculate the worst-case excess risk 
\begin{align}\label{eq:delta-hat-isoDRL}
\hatDcal = &\max_{\Vert w\Vert_{\infty} \leq \wbnd} \quad  \frac{1}{n} \sum_{i \in \calD_3} (w_i-1) r^{\iso}_i \quad {\rm s.t.}\quad \frac{1}{n}\sum_{i \in \calD_3} w_i=1, \;\frac{1}{n}\sum_{i \in \calD_3} w_i \log w_i \leq \rho. 
\end{align}
Given these estimates of the worst-case excess risks, we compare the following methods: (1) DRL: CP interval $C_{1-\tilde \alpha}$, where $\tilde \alpha = \max\{0, \alpha - \hat \Delta(\alpha)\}$.\footnote{To explain this construction, recall from Section~\ref{sec:intro} that we can use the excess risk estimate to choose a tuning parameter that
achieves a desired bound on risk. Specifically, for any value of $\tilde\alpha$, we can bound the risk (i.e., the miscoverage)
for the CP interval $C_{1-\tilde\alpha}$ as $\EE_{\tPtar}[Y\not\in C_{1-\tilde\alpha}(X)] \leq \EE_{\tPtrain}[Y\not\in C_{1-\tilde\alpha}(X)] + \Delta(R_{\tilde\alpha};\calB_{f,\rho}) \leq \tilde\alpha + \Delta(R_{\tilde\alpha};\calB_{f,\rho})$
 (where $R_{\tilde\alpha}$
is the risk defined by the CP interval $C_{1-\tilde\alpha}$, for any value of $\tilde\alpha$). Since $a\mapsto R_a$ is nondecreasing,
this also implies that $a\mapsto \Delta(R_a;\calB_{f,\rho}) $ is nondecreasing (recall from Section~\ref{sec:no-cone}
that $\Du$ is monotone in $R$, as a corollary of Proposition~\ref{prop:ucons-kkt}). Thus, for $\tilde\alpha \leq \alpha$ we have $\EE_{\tPtar}[Y\not\in C_{1-\tilde\alpha}(X)] \leq \tilde\alpha + \Delta(R_{\alpha};\calB_{f,\rho})\approx \tilde\alpha + \hat \Delta(\alpha)$. Consequently, the above
choice of $\tilde\alpha$ ensures that miscoverage will be (approximately) bounded by $\alpha$. A similar argument also holds for iso-DRL-$\wprior$.}; (2) iso-DRL-$\wprior$: CP interval  $C_{1-\alpha_{\iso}}$, where  $\alpha_{\iso} = \max\{0, \alpha - \hatDcal\}$.

\subsection{Synthetic dataset}\label{sec:synthetic}
We start with a synthetic example, in which we fix $n_{\tr} = n = n_{\te} = 500$ and will vary $n_{\rm pre}$ to see how will the initial density ratio estimation $\wprior$ affect the result. We will consider two settings---the ``well-specified'' and ``misspecified'' settings.
Specifically, for the marginal distributions of $X$, we set the well-specified setting with $\Ptrain: X \sim \calN(\mathbf{0}_d,\bI_d)$ and $\Ptar: X \sim \calN(\mu, \bI_d)$,
and the misspecified setting with $\Ptrain: X \sim \calN(\mathbf{0}_d,\bI_d)$ and $\Ptar: X \sim \calN(\mu, \bI_d + \frac{\zeta}{d}\mathbf{1}_d \mathbf{1}_d^\top)$,
where $d = 20$, $\mu = (2/\sqrt{d}) \cdot (1,\cdots,1)^\top$, and $\zeta = 6$.
Since the estimate $\wprior$ for the density ratio will be fitted via logistic regression as described above, 
the first setting is indeed well-specified since, due to the fact that $\Ptrain$ and $\Ptar$ have the same covariance, the logistic model is correct for the distribution
shift from $\Ptrain$ to $\Ptar$. In contrast, the second setting is misspecified since, due to the change in covariance matrix, the 
underlying log-density ratio is no longer a linear function of $\mu^\top X$, and therefore cannot be characterized exactly by a logistic regression model.
Finally, for the conditional distribution of $Y\mid X$, we set $Y \mid X \sim 0.2 \cdot \calN(X^\top \beta + \sin(X_1) + 0.4 X_3^3 + 0.2 X_4^2, 1)$
for both training and target distributions, where $\beta \sim \calN(\mathbf{0}_d,\bI_d)$.

\subsubsection{Results with varying sample size \texorpdfstring{$n_{\rm pre}$}{npre} for estimating \texorpdfstring{$\wprior$}{w0}}
We first consider the scenario with an estimated density ratio $\wprior$.
Recall that we use the subsets $\calD_1 \subset \calD_{\tr}$ and $\calD_{\te,1} \subset \calD_{\te}$ with $|\calD_1| = |\calD_{\te,1}| = n_{\rm pre}$ for estimating $\wprior$;
consequently, for larger values of $n_{\rm pre}$, we will expect a more accurate $\wprior$.
By varying $n_{\rm pre}$ in $\{40,60,80,100,120,140,160\}$, we aim to investigate the robustness of WCP and iso-DRL with respect to the accuracy in $\wprior$, where we fix $\rho = \rho^* \coloneqq D_{\rm KL}(\Ptar \| \Ptrain)$.

\begin{figure}[ht]
    \centering
    \begin{subfigure}[t]{0.32\textwidth}
    \includegraphics[height=0.155\textheight]{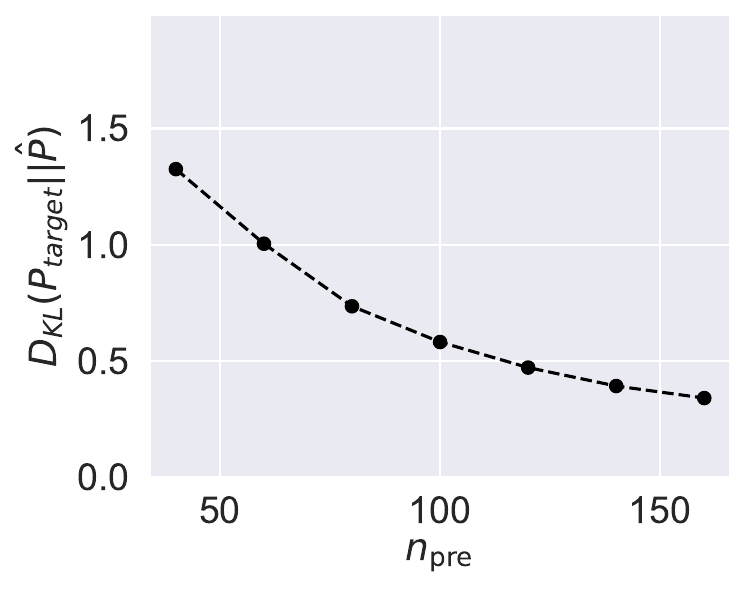}
    \caption{$D_{\rm KL}(\Ptar \| \hat P)$ vs. $n_{\rm pre}$.}
    \label{fig:hat-rho-0}
    \end{subfigure}
    \begin{subfigure}[t]{0.62\textwidth}
        \centering 
        \includegraphics[height=0.16\textheight]{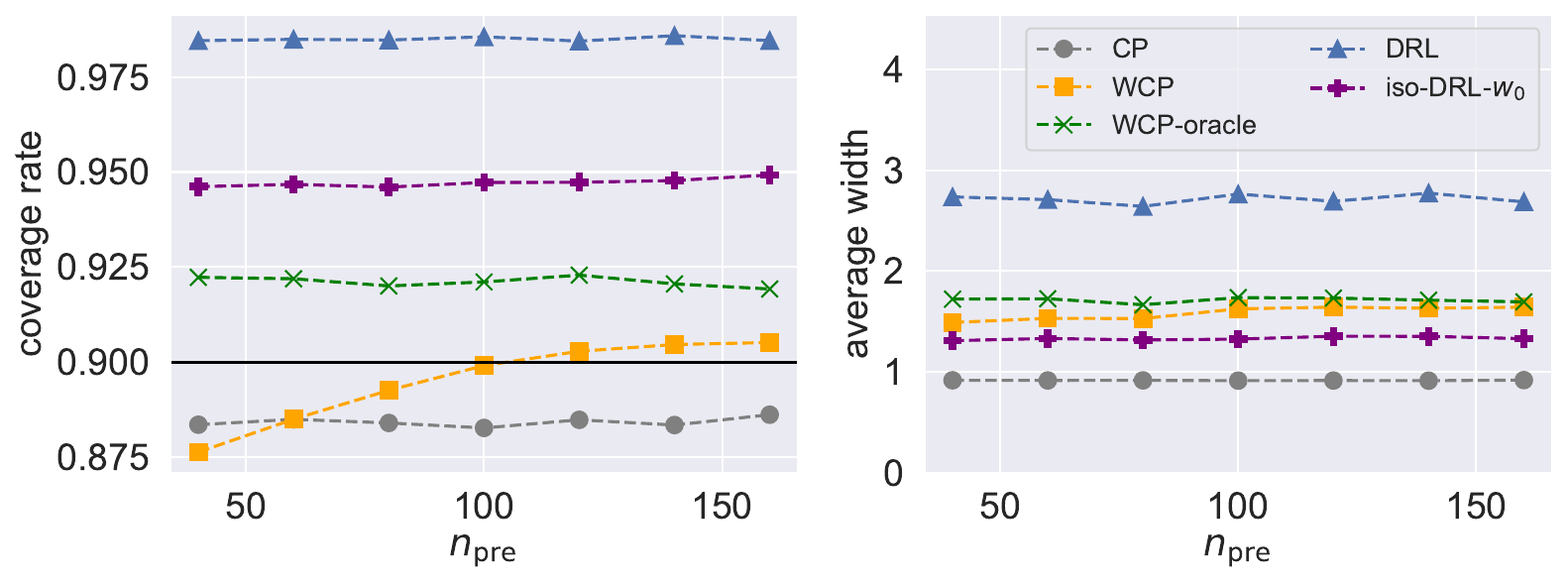}
        \caption{Comparison of all methods with varying $n_{\rm pre}$.} 
        \label{fig:varying-n-0}
    \end{subfigure}
    \caption{Results in the well-specified setting.} 
    \label{fig:simu-0}
\end{figure}

\begin{figure}[ht]   
    \centering 
    \begin{subfigure}[t]{0.32\textwidth}
    \includegraphics[height=0.155\textheight]{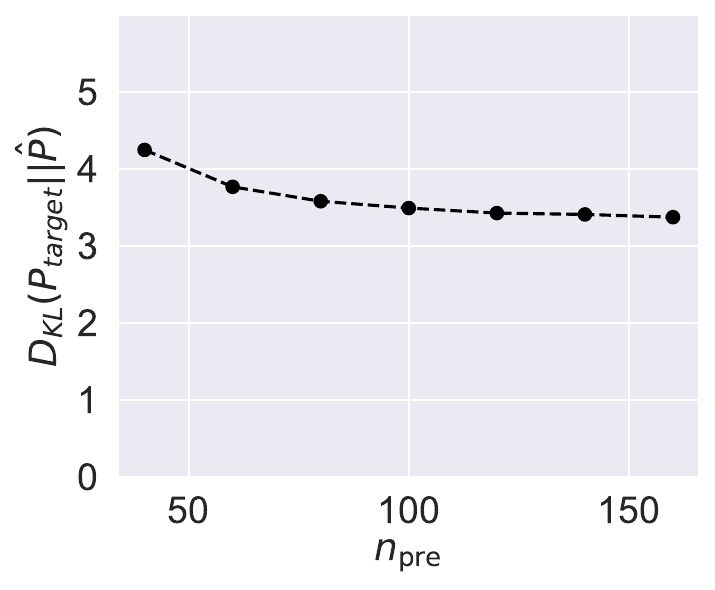}
    \caption{$D_{\rm KL}(\Ptar \| \hat P)$ vs.\ $n_{\rm pre}$.}
    \label{fig:hat-rho-0.25}
    \end{subfigure}
    \begin{subfigure}[t]{0.62\textwidth}
    \includegraphics[height=0.16\textheight]{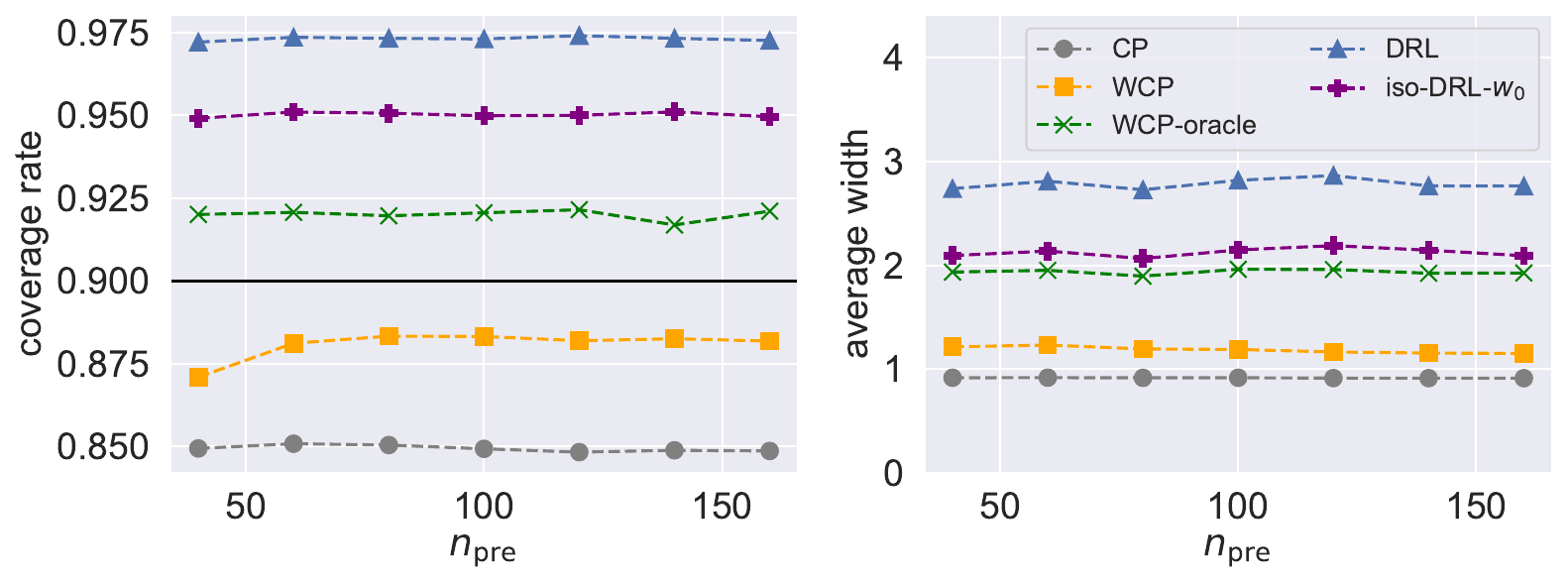}
        \caption{Comparison of all methods with varying $n_{\rm pre}$.} 
    \label{fig:varying-n-0.25}
    \end{subfigure}
    \caption{Results in the misspecified setting.}
    \label{fig:simu-0.25}
\end{figure}

\paragraph{Well-specified setting.}
In Figure~\ref{fig:varying-n-0}, where the solid horizontal line (in the middle plot) marks the nominal coverage level, $1-\alpha = 90\%$, we can see that the uncorrected CP exhibits undercoverage due to the mismatch between $\Ptar$ and $\Ptrain$, while the coverage of WCP using $\wprior$ increases to $90\%$ as $n_{\rm pre}$ increases, since $\wprior$ becomes more accurate with larger $n_{\rm pre}$ (cf.~Figure~\ref{fig:hat-rho-0}). The generic DRL, even with $\rho = \rho^*$, tends to be conservative and has the widest interval. In comparison, iso-DRL-$\wprior$ has coverage very close to the target level.

\paragraph{Misspecified $\wprior$. }
In Figure~\ref{fig:varying-n-0.25}, we show results for the misspecified setting.
Since $\wprior$ is estimated from a model class that does not contain the true density ratio,
consequently $D_{\rm KL}(\Ptar \| \hat P)$ does not converge to zero as $n_{\rm pre}$ increases (cf.~Figure~\ref{fig:hat-rho-0.25}). As a result, both uncorrected CP and WCP (which is weighted
with the misspecified $\wprior$) exhibit undercoverage. The proposed iso-DRL-$\wprior$ method has coverage slightly above $90\%$ but has interval width close to that of WCP-oracle, while DRL is overly conservative.

\begin{figure}[ht]   
    \centering 
    \begin{subfigure}[t]{0.49\textwidth}
    \includegraphics[height=0.175\textheight]{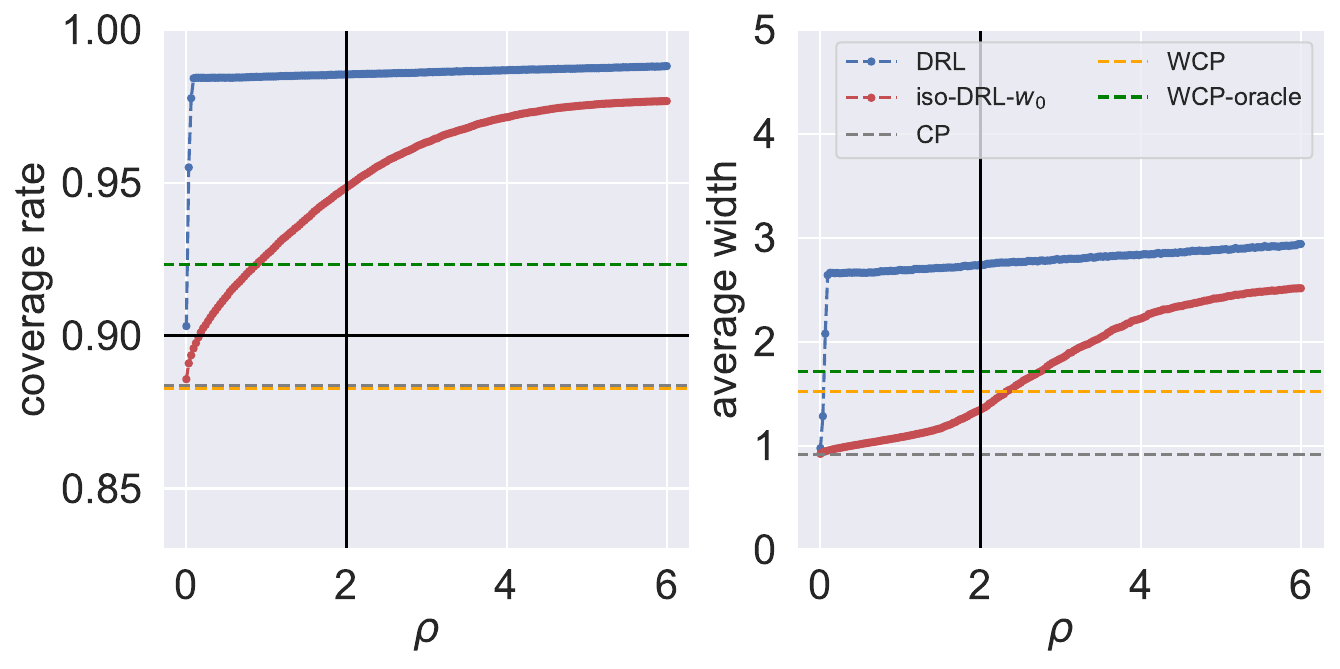}
    \caption{Results in the well-specified setting.}
    \label{fig:generic_1}
    \end{subfigure}
    \begin{subfigure}[t]{0.49\textwidth}
    \includegraphics[height=0.175\textheight]{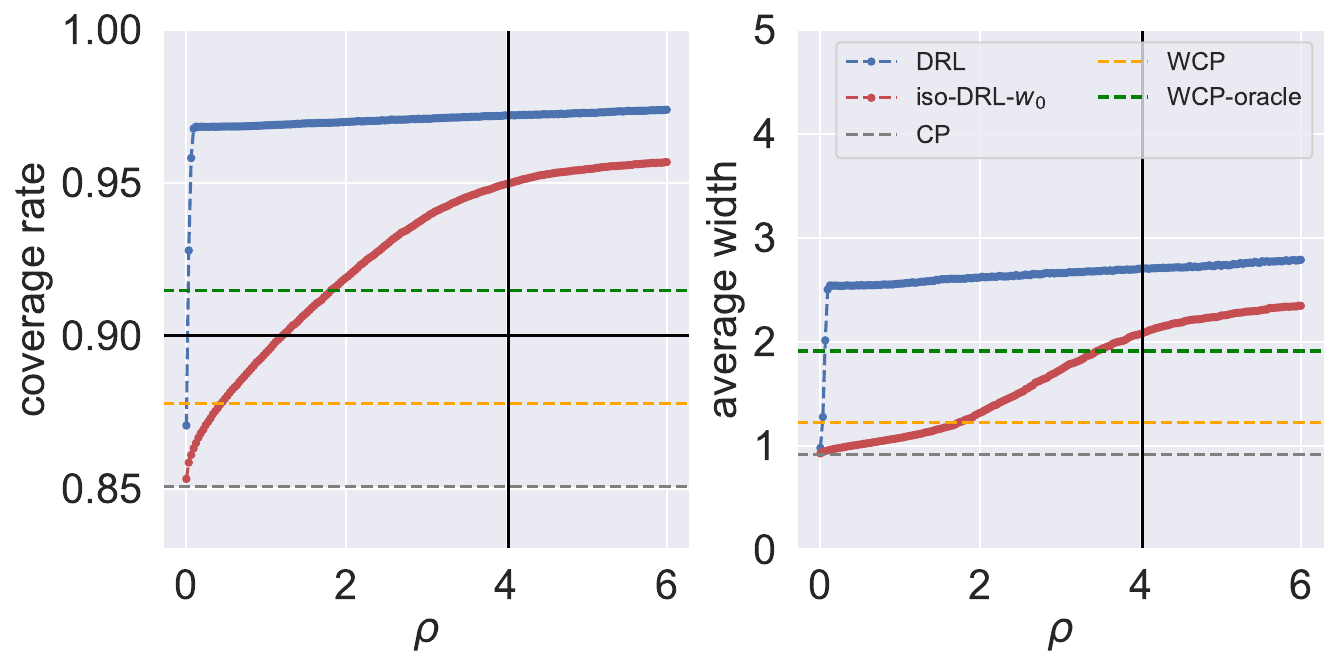}
        \caption{Results in the misspecified setting.} 
    \label{fig:generic_2}
    \end{subfigure}
    \caption{Results with varying $\rho$.}
    \label{fig:rho-v}
\end{figure}



\subsubsection{Results with varying \texorpdfstring{$\rho$}{rho}}
In this section, we investigate the sensitivity of each approach (DRL and iso-DRL-$\wprior$) to the choice of $\rho$. 
Fixing $n_{\rm pre} = 50$, we vary $\rho$ in $[0.002, 6]$. The solid vertical line in each plot denotes the true KL divergence, $\rho^* = D_{\rm KL}(\Ptar \| \Ptrain)$.  
Other methods that do not depend on $\rho$ behave in the same way as shown in the previous section.

We can see from both plots that the prediction intervals produced by DRL are quite conservative and much wider than the oracle interval across nearly the entire range of $\rho$, even values $\rho$ much smaller than the true distribution shift magnitude $\rho^* = D_{\rm KL}(\Ptar \| \Ptrain)$. In comparison, 
for iso-DRL-$\wprior$, when $\rho=\rho^*$, the width of intervals is comparable to the oracle interval in both cases, and the coverage and width vary slowly as we change the value of $\rho$. From this we can see that the isotonic constraint offers a significant gain in accuracy if we have a reasonable estimate of $\rho^*$.

\subsection{Real data: \texttt{wine quality} dataset}\label{sec:real-simu}
\begin{figure}[htb]
    \centering
    \begin{subfigure}[t]{0.29\textwidth}
    \hspace*{-0.25in}
    \includegraphics[height=0.19\textheight]{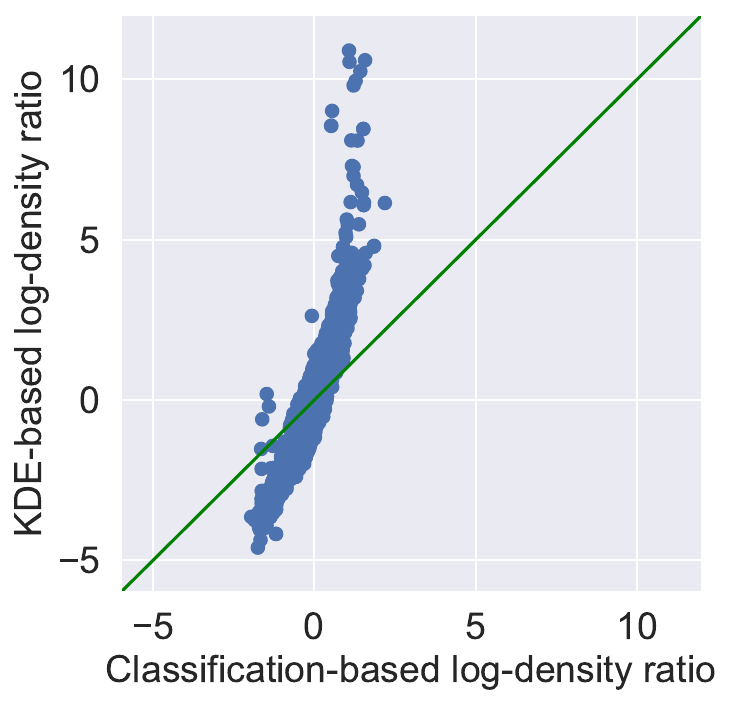}
    \caption{Log-density ratio estimation: KDE versus logistic regression.}
    \label{fig:wine-log-density-ratio}
    \end{subfigure}
    \begin{subfigure}[t]{0.69\textwidth}
    \includegraphics[height=0.19\textheight]{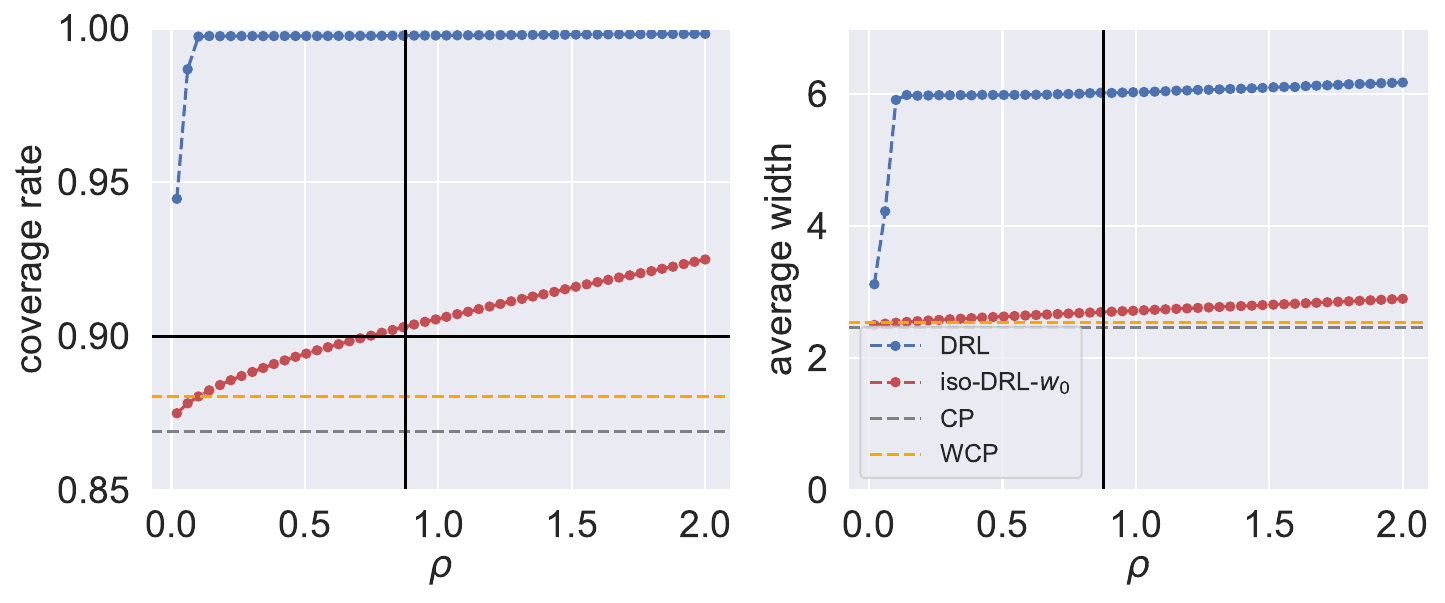}
    \caption{Comparison with varying $\rho$ (sample size $n_{\rm pre} = 50$).}
    \label{fig:wine-varying-rho}
    \end{subfigure}
    \caption{Results for \texttt{wine quality} dataset.}
    \label{fig:simu-wine}
\end{figure}

We next consider a real dataset: the \texttt{wine quality} dataset  \citep{cortez2009modeling}\footnote{Available at \url{https://archive.ics.uci.edu/dataset/186/wine+quality}}. The dataset includes $12$ variables that measure the physicochemical properties of wine and we treat the variable \texttt{quality} as the response of interest. The entire dataset consists of two groups: the white and red variants of the Portuguese ``Vinho Verde'' wine ($1599$ data points for the red wine and $4898$ data points for the white wine). The subset of red wine is treated as the test dataset and that of white wine is viewed as the training set. All variables are nonnegative and we scale each variable by its largest value such that the entries are bounded by $1$. Similar to the dataset partition in synthetic simulation, we fix $n_{\rm pre} = 50$, $n_{\tr} = n = 1900$, and $n_{\te} = 1000$. We first fit a kernel density estimator (Gaussian kernel with a bandwidth suggested by cross-validation) using the entire dataset as a proxy of the oracle density ratio. 
Figure~\ref{fig:wine-log-density-ratio} plots this against the 
log-density ratio obtained from logistic regression fitted on $n_{\rm pre}$ samples from each group. 
It can be seen that the two density ratios exhibit an approximately isotonic trend, which motivates us to consider the isotonic constraint with respect to $\wprior$.

To assess the performance of the proposed approach, we estimate $\wprior$ using the same procedure as for the simulated data, with sample size $n_{\rm pre} = 50$ for each group. 
We consider the uncertainty set defined by KL-divergence and choose $\rho$ from $50$ uniformly located grid points in $[0.02, 2]$ in Figure~\ref{fig:wine-varying-rho}. The solid vertical line (in the middle and right plots) denotes an estimate $\widehat\rho$ of the KL divergence $D_{\rm KL}(\Ptar \| \Ptrain)$ to ensure that we are considering a reasonable range of values of $\rho$ (see Appendix~\ref{sec:add-simu_wine} for details on this estimate). 
In Figure~\ref{fig:wine-varying-rho}, similar to the performance in Section~\ref{sec:synthetic} for simulated data, DRL tends to be conservative: the coverage rate quickly approaches $1$ while $\rho$ is still below $0.1$ and the intervals tend to be wide. In the meantime, iso-DRL-$\wprior$ captures the approximate isotonic trend in Figure~\ref{fig:wine-log-density-ratio} and achieves valid coverage by recalibrating the weighted approach. The key message is that in the real data case, even when there is no oracle information for selecting $\rho$ and the isotonic trend is not exact, the proposed iso-DRL-$\wprior$ with the isotonic constraint with respect to the pre-fitted density ratio is less sensitive to the selection of $\rho$.

\section{Additional related work}\label{sec:related}
In this section, we discuss some additional literature in several related areas, including transfer learning, DRL, sensitivity analysis, shape-constrained learning, and conformal prediction.  
\paragraph{Transfer learning.} 
Transfer learning, in which data from one distribution is used to improve performance on a related but different distribution, is usually categorized into domain adaptation and inductive transfer learning  \citep{Redko2020ASO}.

Domain adaptation focuses on the scenario with covariate shift.
From the theoretical side, the performance of machine learning models is analyzed in \cite{BenDavid2010ImpossibilityTF,BenDavid2012OnTH,pathak2022new, pathak2024design,Hanneke2019OnTV}.  
To implement efficient predictions, weighted methods are adopted as the first trial to draw $P$ closer to $Q$ \citep{Cortes2008SampleSB,Gretton2009CovariateSB, ma2023optimally, ge2023maximum}. Another scenario requires a small number of labeled target samples, which can be feasible in reality and related works include \cite{Chen2011CoTrainingFD,Chattopadhyay2013JointTA,Yang2012ATO}, etc.
Inductive transfer learning, on the other hand, assumes that the marginal distributions of $X$ are invariant for training and target distributions and is studied in difference statistical settings \citep{Bastani2021PredictingWP,Cai2019TransferLF,Li2021TransferLF,Tian2021TransferLU}.

\paragraph{Distributionally robust learning (DRL).} 
Our work is directly related to DRL~\citep{ben1998robust,el1997robust}, 
which aims to control certain statistical risks uniformly over a set of candidate distributions for the target distribution. 
Different classes of the uncertainty set are studied in the literature, such as the optimal transport discrepancy \citep{shafieezadeh2015distributionally, blanchet2019quantifying, blanchet2019robust, esfahani2015data} and $f$-divergence \citep{duchi2021statistics, duchi2018learning,weiss2023bilateral}. 
Further constraints on the uncertainty set as the improvement of DRL are explored by \cite{duchi2019distributionally, setlur2023bitrate,esteban2022partition,liu2023geometry,popescu2007robust,shapiro2023conditional}. 
The recent work of \cite{wang2023distributionally} considers the constraint that the unseen target distribution is a mixture of data distributions from multiple sources.

\paragraph{Sensitivity analysis.} Sensitivity analysis is closely related to DRL and is widely studied in the field of causal inference \citep{cornfield1959smoking, rosenbaum1987sensitivity, tan2006distributional, ding2016sensitivity, zhao2019sensitivity,de2023hidden} with the goal of evaluating the effect of unmeasured confounders and relaxing untestable assumptions. Sensitivity models can be viewed as a specific example of constraints on distribution shift. 
For example, 
the marginal $\Gamma$-selection model \citep{tan2006distributional} 
with a binary treatment $T$ imposes a bound constraint on the distribution shift from the data distribution $P_{Y(1) \mid X, T=1}$ to the counterfactual $P_{Y(1) \mid X, T=0}$. Recent works also investigate 
the sensitivity model from the perspective of DRL, such as \cite{yadlowsky2018bounds, jin2022sensitivity, jin2023sensitivity, sahoo2022learning}. Sensitivity analysis that incorporates more informative constraints is explored in \cite{huang2024variance,nie2021covariate}. 

\paragraph{Statistical learning with shape constraints.} Our work also borrows ideas from shape-constrained learning,
which have been studied across various applications \citep{grenander39theory,matzkin1991semiparametric}. The isotonic constraint is the most common one among these. 
Since \cite{rao1969estimation}, the properties of isotonic regression are well studied in the literature \citep{brunk1957minimizing,brunk1972statistical,zhang2002risk,han2019isotonic,yang2019contraction,durot2018limit}. Moreover, the isotonic constraint is also widely applied to calibration for distributions in regression and classification settings \citep{zadrozny2002transforming,niculescu2012obtaining,van2023causal,henzi2021isotonic,berta2024classifier}.

\paragraph{Conformal prediction.} 
One important application of iso-DRL is to recalibrate conformal prediction intervals. 
Conformal prediction \citep{vovk2005algorithmic,shafer2008tutorial} provides a framework for distribution-free uncertainty quantification, which constructs prediction intervals that are valid with exchangeable data from any underlying distribution and with any ``black-box'' algorithm. 
As the validity of WCP \citep{Tibshirani2019ConformalPU} with the estimated density ratio only holds up to a coverage gap due to the error the estimate $\wprior$  \citep{lei2020conformal,candes2023conformalized,gui2024conformalized}, the work \cite{jin2023sensitivity} further establish a robust guarantee via sensitivity analysis. Besides the weighted approaches, there are other solutions in the literature: \cite{cauchois2020robust,ai2024not} address the issue of joint distribution shift via the DRL; \cite{qiu2023prediction,yang2024doubly,chen2024biased} formulate the covariate shift problem within the semiparametric/nonparametric framework and utilize the doubly-robust theory to correct the distributional bias.

\section{Discussion}\label{sec:discuss}
In this paper, we focus on distributionally robust risk evaluation with the isotonic constraint on the density ratio as the regularization, which aims to avoid over-pessimistic candidate distributions. 
This is similar in flavor to many tools in high-dimensional statistical learning, where regularization/inductive bias is introduced to improve generalization. We provide an efficient approach to solve the shape-constrained optimization problem via an equivalent reformulation, for which estimation error bounds for the worst-case excess risk are also provided.
To conclude, we provide further discussions on the proposed iso-DRL framework and highlight several open questions.


\paragraph{Stability against distribution shift.}
Excess risk can also be interpreted from the perspective of stability against distribution shift \citep{lam2016robust,namkoong2022minimax,rothenhausler2023distributionally}. With a fixed budget $\vareps \ll 1$ for the excess risk, the largest tolerance of distribution shift such that the excess risk is under control is of interest. Taking the $f$-divergence constraint as an example, to ensure $\Df \leq \vareps \ll 1$, then $\rho$ needs to obey
$\rho \leq (2 \var(R(X)))^{-1}f^{\prime\prime}(1) \cdot \vareps^2 + o(\vareps^2)$ \citep{lam2016robust, duchi2018learning, blanchet2023statistical}.
However, with the additional isotonic constraint on the density ratio, we can tolerate larger distribution shift:
$\rho \leq (2 \var([\pi(R)](X)))^{-1}f^{\prime\prime}(1) \cdot \vareps^2 + o(\vareps^2)$.
This improvement implies that when side information of the underlying distribution shift is provided, risk evaluation will be less sensitive to the hyperparameters describing the uncertainty set (e.g., $\rho$), thus is more robust with the presence of distribution shift.

\paragraph{From risk evaluation to distributionally robust optimization.}
Different from risk evaluation, distributionally robust optimization (DRO) focuses on the optimization problem with a loss function $\ell_{\theta}(x)$, i.e.,
$\hat \theta \in \argmin {\theta \in \Theta} \sup_{Q \in \calQ} \EE_Q \;\ell_{\theta}(X)$.
Under smoothness conditions on $\ell_{\theta}$, asymptotic normality for $\hat \theta$ is established in the literature \citep{duchi2018learning}. The DRO framework is shown to regularize $\hat \theta$ in terms of variance penalization \citep{lam2016robust,duchi2018learning} or explicit norm regularization \citep{blanchet2019quantifying}. It is interesting to incorporate the isotonic constraint into DRO and to understand the effect of the isotonic constraint in the asymptotics of $\hat \theta^{\iso}$.


\section*{Acknowledgements}
R.F.B. was supported by the Office of Naval Research via grant N00014-20-1-2337, and by the National Science Foundation via grant DMS-2023109. C.M. was partially supported by the National Science Foundation via grant DMS-2311127.

\section*{Appendix}

\appendix

\section{Proofs of results in Section~\ref{sec:setup}}\label{sec:proof-setup}

\subsection{Proof of Proposition~\ref{prop:ucons-kkt}}\label{sec:proof-prop-kkt}

It is straightforward to check that $\Du$ is always an upper bound of the new formulation stated in 
Proposition~\ref{prop:ucons-kkt}, simply by taking $w = \phi\circ R$. 
Therefore, it remains to show the converse: under Condition~\ref{cond:B-cond}, $\Du$ is also a \emph{lower} bound of the new formulation stated in Proposition~\ref{prop:ucons-kkt}. 

To this end, it suffices to prove that for any $w_{\myhash} \Ptrain \in \calB$, there exists a nondecreasing function $\phi$ such that $(\phi \circ R)_{\myhash} \Ptrain \in \calB$, and 
\[
\EE_{\Ptrain}\left[w(X)R(X)\right] \leq \EE_{\Ptrain}\left[\phi(R(X))R(X)\right].
\]
We construct such a function $\phi$ in two steps. 

\paragraph{Step 1: Conditioning. }
For any $w$ such that $w_{\myhash} \Ptrain \in \calB$, we define $g$ as a measurable function satisfying 
\[
g(R(X)) = \EE\left[w(X) \mid R(X)\right],\quad \Ptrain\text{-almost~surely}.
\]
(Note that $g$ is not necessarily a monotone function.)
As a result, by the tower law, we have 
\begin{align}\label{eq:conditioning}
\EE_{\Ptrain}[w(X)R(X)] = \EE_{\Ptrain}[g(R(X)) R(X)].
\end{align}
Since $w_{\myhash} \Ptrain \in \calB$, by Jensen's inequality, for any convex function $\psi$, we have 
\[
\EE_{\Ptrain}\left[\psi\left(g(R(X))\right)\right] = \EE_{\Ptrain}\left[\psi\left(\EE\left[w(X) \mid R(X)\right]\right)\right] \leq \EE_{\Ptrain}\left[\psi(w(X))\right],
\]
which implies $(g \circ R)_{\myhash} \Ptrain \in \calB$ by Condition~\ref{cond:B-cond}.

\paragraph{Step 2: Rearrangement. }
Denote $F_1$ and $F_2$ as the cumulative distribution functions of $g(R(X))$ and $R(X)$, respectively. Let $U \sim {\rm Unif}([0,1])$. Then, we have $F_1^{-1}(U) \overset{d}{=} g(R(X))$ and $F_2^{-1}(U) \overset{d}{=} R(X)$, where $F_k^{-1}$ is the generalized inverse of $F_k$ for $k = 1, 2$,
and where $\overset{d}{=}$ denotes equality in distribution. Moreover, $F_1^{-1}$ is nondecreasing and
\[
g(F_2^{-1}(U)) \overset{d}{=} g(R(X)) \overset{d}{=} F_1^{-1}(U),
\]
which implies that $F_1^{-1}$ is the monotone rearrangement of $g \circ F_2^{-1}$.
By \cite[eqn.~(378)]{hardy1952inequalities}, we have
\begin{align}\label{eq:rearrangement}
\EE_{\Ptrain}\left[g(R(X)) R(X)\right] = \EE\left[g(F_2^{-1}(U)) F_2^{-1}(U)\right] \leq \EE\left[F_1^{-1}(U) F_2^{-1}(U)\right].
\end{align}

Next, let $\phi$ be a measurable function satisfying
\[\phi(F_2^{-1}(U)) = \EE\left[F_1^{-1}(U) \mid F_2^{-1}(U)\right],\]
almost surely with respect to the distribution $U\sim\textnormal{Unif}([0,1])$. 
Since $F_k^{-1}$ is the generalized inverse of a CDF $F_k$, for each $k=1,2$, it is therefore monotone 
nondecreasing. Therefore, we can choose $\phi$ to be a monotone nondecreasing function.
Moreover, to verify that $(\phi \circ R)_{\myhash} \Ptrain \in \calB$, we will check that $\phi(R(X)) \overset{cvx}{\orderless} g(R(X))$ (and use 
Condition~\ref{cond:B-cond}, along with the fact that $(g\circ R)_{\myhash}\Ptrain\in\calB$ as established above): for any convex
function $\psi$, we have
\begin{align*}
\EE_{\Ptrain}[\psi(\phi(R(X)))] &\overset{d}{=}
\EE[\psi(\phi(F_2^{-1}(U)))]\\ &= \EE[\psi(\EE\left[F_1^{-1}(U) \mid F_2^{-1}(U)\right]
)] \leq  \EE[\psi(F_1^{-1}(U))] = \EE_{\Ptrain}[\psi(g(R(X)))],
\end{align*}
where the inequality holds by Jensen's inequality.

We then have
\begin{align*}
\EE\left[F_1^{-1}(U) F_2^{-1}(U)\right] & = \EE\left[\EE\left[F_1^{-1}(U) \mid F_2^{-1}(U) \right] F_2^{-1}(U)\right]\\* 
& = \EE\left[\phi(F_2^{-1}(U))F_2^{-1}(U)\right] = \EE_{\Ptrain}\left[\phi(R(X))R(X)\right].
\end{align*}
This equality, combined with~\eqref{eq:conditioning} and~\eqref{eq:rearrangement}, yields the desired outcome: 
$\EE_{\Ptrain}\left[w(X)R(X)\right] \leq \EE_{\Ptrain}\left[\phi(R(X))R(X)\right]$. We hence complete the proof.

\section{Proofs  of results in Section~\ref{sec:iso}}\label{sec:proof-iso}

\subsection{Preliminaries}
Before we present the proof, we begin with some preliminaries: we introduce some notation,
definitions, and facts that will aid in the proof below.

\subsubsection{Adding an \texorpdfstring{$L_2$}{l2} constraint}
First, we will define a version of our optimization problem that defines $\Du$, by adding an $L_2$ constraint:
\begin{align}\label{opt:l2}
\Dul \quad = \quad \sup_{w \geq 0,\;w \in L_2(\Ptrain)} \qquad & \EE_{\Ptrain}\left[w(X)R(X)\right] - \alphaP  \nonumber\\*
{\rm subject~to} \qquad & w_{\myhash} \Ptrain\in \calB.
\end{align}
We can observe that, by construction,
\[\Dul = \Delta(R;\calB \cap \calB_{L_2}),\]
where $\calB_{L_2}$ is the set of all distributions with finite second moment.
The following result verifies that adding the $L_2$ constraint does not change the outcome of the optimization problem:
\begin{proposition}\label{prop:l2-approx}
Under the notation and definitions above, it holds that $\Du = \Dul$.
\end{proposition}
\noindent We defer the proof of this proposition to Section~\ref{sec:proof-l2}.

\subsubsection{The isotonic projection}
We next review some facts regarding the isotonic projection operator $\pi$. 
To ease notation, we denote $\la a, b \ra_{\Ptrain} = \int_{\calX} a(x) b(x) \mathsf{d}\Ptrain(x)$ for any functions $a,b\in L_2(\Ptrain)$.
 
The first property relates to the isotonic projection as a projection to a convex cone (\cite{bauschke2019convex}, Theorem 3.14; \cite{edwards2012functional}, Proposition 1.12.4):
\begin{equation}\label{eqn:iso_fact_2}
\textnormal{For any $w\in L_2(\Ptrain)$ and any $v\in \Ciso \cap L_2(\Ptrain)$, $\langle v, w - \pi(w)\rangle_\Ptrain \leq 0$.}
\end{equation}
Moreover, it holds that (\cite{brunk1963extension}, Theorem 1; \cite{brunk1965conditional}, Corollary 3.1):
\begin{equation}\label{eqn:iso_fact_3}
\textnormal{For any $w\in L_2(\Ptrain)$ and any $h:\mathbb{R}\to\mathbb{R}$, $\langle h\circ \pi(w), w - \pi(w)\rangle_\Ptrain = 0$.}
\end{equation}
In particular, by chosing $h(t)\equiv 1$, we can see that isotonic projection preserves the mean,
\begin{equation}\label{eqn:iso_fact_1}
\textnormal{For any $w\in L_2(\Ptrain)$, \;$\EE_{\Ptrain}[w(X)] = \EE_{\Ptrain}[[\pi(w)](X)]$.}
\end{equation}
Finally, we relate the isotonic projection to the convex ordering:
\begin{equation}\label{eqn:iso_fact_4}
\textnormal{For any $w\in L_2(\Ptrain)$, $\pi(w)\overset{cvx}{\orderless} w$.}
\end{equation}

To see \eqref{eqn:iso_fact_4}, for any convex function $\psi$, by the nonnegativity of Bregman divergence \citep{bregman1967relaxation}, it holds that
\begin{align*}
\la \psi(w) - \psi(\pi(w)), 1 \ra_{\Ptrain} \geq \la \psi^\prime \circ \pi(w), w - \pi(w) \ra_{\Ptrain}.
\end{align*}
According to the property \eqref{eqn:iso_fact_3}, we further obtain $\la \psi^\prime \circ \pi(w), w - \pi(w) \ra_{\Ptrain}$, which implies that $\pi(w)\overset{cvx}{\orderless} w$ by definition.

\subsection{Proof of Theorem~\ref{thm:equiv}}\label{sec:proof-equiv}
We split the proof into three steps: 
\begin{enumerate}
\item prove that $\Diso \leq \DRiso$;
\item prove that $\Diso = \DRiso$ provided that Condition~\ref{cond:B-cond} holds;
\item prove the claim on attainability of minimizers provided that Condition~\ref{cond:B-cond} holds.
\end{enumerate}

\paragraph{Step 1: Prove $\Diso \leq \DRiso$.}
By the definition of $\Diso$ as the supremum in the optimization problem~\eqref{eq:opt-iso0}, 
for any $\vareps > 0$, there exists a feasible $w_{\vareps}$ such that
\begin{align}\label{eq:eps-original}
\EE_{\Ptrain}[w_{\vareps}(X) \cdot R(X)] - \alphaP \geq  \Diso - \vareps.
\end{align}
Next, define a sequence of truncated functions, $w_{\vareps,n}(x) = \min\{w_{\vareps}(x), n\}$. 
Since $w_{\vareps}\in\Ciso$, it holds that $w_{\vareps,n}\in\Ciso$ as well, and moreover since the truncated
function is bounded we also have  $w_{\vareps,n}\in L_2(P)$. By fact~\eqref{eqn:iso_fact_2}, it therefore holds that
\[
\EE_{\Ptrain}[w_{\vareps,n}(X) \cdot (R - [\pi(R)])(X)] = \la w_{\vareps,n}, R - \pi(R) \ra_{\Ptrain} \leq 0,
\]
for each $n\geq 1$. Then, by the dominated convergence theorem, taking a limit as $n\to\infty$ we obtain
\begin{align}\label{eq:eps-nonpos}
\EE_{\Ptrain}[w_{\vareps}(X) \cdot (R - [\pi(R)])(X)]  \leq 0.
\end{align}
Moreover, $\EE_{\Ptrain}[[\pi(R)](X)] = \alphaP$ by~\eqref{eqn:iso_fact_1}. Combining everything, then,
\begin{align*}
\Diso - \vareps & \leq  \EE_{\Ptrain}[w_{\vareps}(X) \cdot R(X)]  - \alphaP\\ &\leq  \EE_{\Ptrain}[w_{\vareps}(X) \cdot [\pi(R)](X)] - \EE_{\Ptrain}[[\pi(R)](X)] \leq \DRiso,
\end{align*}
where the last step holds since, because $w_{\vareps}$ is feasible for the optimization problem~\eqref{eq:opt-iso0} that defines $\Diso$,
it is also feasible for $\DRiso$ (i.e., $w\geq 0$ and $w_{\myhash}\Ptrain\in\calB$).
Since $\vareps > 0$ is arbitrary, we obtain the desired result $\Diso \leq \DRiso$. 

\paragraph{Step 2: Prove $\Diso = \DRiso$ under Condition~\ref{cond:B-cond}.} 
By Proposition~\ref{prop:l2-approx}, we have $\DRiso =  \Delta_2(\pi(R);\calB) = \Delta(\pi(R);\calB\cap \calB_{L_2})$.
Next, note that if Condition~\ref{cond:B-cond} holds for $\calB$, then this condition holds for $\calB\cap \calB_{L_2}$ as well
(because for any $Q^{\prime} \overset{cvx}{\orderless} Q$, we have $\EE_{Q'}[X^2] \leq \EE_Q[X^2]$ by definition
of the convex ordering---and so if $Q\in \calB_{L_2}$ then $Q'\in \calB_{L_2}$ as well.)
Therefore, we can apply  Proposition~\ref{prop:ucons-kkt}  to the term $\Delta(\pi(R);\calB\cap \calB_{L_2})$, which yields
 the following equivalent formulation:
\begin{align}\label{opt:l2-}
\Delta(\pi(R);\calB\cap \calB_{L_2}) \quad = \quad \sup_{\phi: \RR \rightarrow \RR_+} \qquad & \EE_{\Ptrain}\left[(\phi \circ \pi(R))(X)\cdot [\pi(R)](X)\right] - \EE_{\Ptrain}[[\pi(R)](X)]   \nonumber\\*
{\rm subject~to} \qquad & (\phi \circ \pi(R))_{\myhash} \Ptrain\in \calB, \quad \phi \circ \pi(R) \in L_2(\Ptrain), \quad \phi \text{~is~nondecreasing}.
\end{align}
Then, for any $\vareps>0$, there exists some $\phi_{\vareps}$ satisfying the above constraints so that
\[ \EE_{\Ptrain}\left[(\phi_{\vareps} \circ \pi(R))(X)\cdot [\pi(R)](X)\right] - \EE_{\Ptrain}[[\pi(R)](X)]    \geq \Delta(\pi(R);\calB\cap \calB_{L_2}) - \vareps= \DRiso - \vareps.\]
Now define $\tilde w_{\vareps} = \phi_{\vareps}\circ \pi(R)$, i.e., we have
\[ \EE_{\Ptrain}\left[\tilde w_{\vareps} (X)\cdot [\pi(R)](X)\right] - \EE_{\Ptrain}[[\pi(R)](X)]    \geq \DRiso- \vareps,\]
where $(\tilde w_{\vareps})_{\myhash}\Ptrain\in\calB$ and $\tilde w_{\vareps}\in L_2(\Ptrain)$, and also $\tilde w_{\vareps}\in\Ciso$, by construction and by feasibility of $\phi_{\vareps}$. 
Moreover, by the facts~\eqref{eqn:iso_fact_1} and~\eqref{eqn:iso_fact_3},
\[\EE_{\Ptrain}[[\pi(R)](X)]=\alphaP, \quad \la \tilde w_{\vareps}, R - \pi(R)\ra_{\Ptrain}
= \la \phi_{\vareps}\circ \pi(R), R - \pi(R)\ra_{\Ptrain}=0,\]
and therefore,
\[ \EE_{\Ptrain}\left[\tilde w_{\vareps} (X)\cdot R(X)\right] - \alphaP    \geq \DRiso- \vareps .\]
But we have verified above that $\tilde w_{\vareps}$ is feasible for the optimization problem~\eqref{eq:opt-iso0} defining $\Diso$, i.e.,
\[\EE_{\Ptrain}\left[\tilde w_{\vareps} (X)\cdot R(X)\right] - \alphaP \leq \Diso.\]
Since $\vareps>0$ is arbitrary, this verifies that $\DRiso\leq \Diso$, and thus completes this step.

\paragraph{Step 3: attainability of minimizers under Condition~\ref{cond:B-cond}. }

Suppose $\DRiso$ is attained at $\tilde w$, i.e.,  
\[
 \EE_{\Ptrain}[\tilde w(X)\cdot [\pi(R)](X)] 
- \EE_{\Ptrain}[[\pi(R)](X)] = \DRiso.\]
By Proposition~\ref{prop:ucons-kkt}, 
we can construct some nondecreasing function $\tilde\phi$, with $(\tilde\phi\circ \pi(R))_{\myhash}\Ptrain\in\calB$, 
such that
\[\DRiso =  \EE_{\Ptrain}[\tilde \phi([\pi(R)](X))\cdot [\pi(R)](X)] - \EE_{\Ptrain}[[\pi(R)](X)].\]
Recalling that $\EE_{\Ptrain}[[\pi(R)](X)]=\alphaP$ by~\eqref{eqn:iso_fact_1},
and $\DRiso=\Diso$ by Steps 1 and 2, we now have
\[\Diso =  \EE_{\Ptrain}[\tilde \phi([\pi(R)](X))\cdot [\pi(R)](X)] - \alphaP.\]
Next, by fact~\eqref{eqn:iso_fact_3},
\[ \EE_{\Ptrain}[\tilde \phi([\pi(R)](X))\cdot (R(X) - [\pi(R)](X) )]
= \la \tilde\phi \circ \pi(R), R-\pi(R)\ra_{\Ptrain} = 0,\]
and so
\[\Diso =  \EE_{\Ptrain}[\tilde \phi([\pi(R)](X))\cdot R(X)] - \alphaP.\]
Therefore, $\Diso$ is attained at $\tilde\phi \circ \pi(R)$ (which, by construction, satisfies $\tilde\phi \circ \pi(R)\in\Ciso$,
as well as $(\phi\circ \pi(R))_{\myhash}\Ptrain\in\calB$ as above, and is therefore feasible).

Conversely, suppose that
$\Diso$ is attained at $\tilde w$, i.e.,
\[
 \EE_{\Ptrain}[\tilde w(X)\cdot R(X)] 
- \alphaP = \Diso.\]
Again applying~\eqref{eqn:iso_fact_1}, and the fact that  $\DRiso=\Diso$ by Steps 1 and 2,
\[\DRiso = 
 \EE_{\Ptrain}[\tilde w(X)\cdot R(X)] 
-   \EE_{\Ptrain}[[\pi(R)](X)] \leq  \EE_{\Ptrain}[\tilde w(X)\cdot [\pi(R)](X)] 
-   \EE_{\Ptrain}[[\pi(R)](X)] ,\]
where for the last step,
since $\tilde w\in\Ciso$ (because it is feasible for $\Diso$), we have
\[ \EE_{\Ptrain}[\tilde w(X)\cdot (R(X) - [\pi(R)](X) )]
= \la \tilde w, R-\pi(R)\ra_{\Ptrain} \leq 0,\]
by~\eqref{eqn:iso_fact_2}.
But $\tilde w$ is feasible for $\DRiso$ (since it is feasible for $\Diso$), and therefore,
we also have
\[\DRiso = \Diso \leq
  \EE_{\Ptrain}[\tilde w(X)\cdot [\pi(R)](X)] 
-   \EE_{\Ptrain}[[\pi(R)](X)] .\]
In other words, $\DRiso$ is attained at $\tilde w$, which completes the proof.

\subsection{Proof of Proposition~\ref{prop:equi-rewt}}\label{sec:proof_prop:equi-rewt}
We formally define $\Disow$ as follows:
\begin{align}\label{eq:wt-delta}
\Disow \quad = \quad \sup_{w \geq 0} \qquad & \EE_{\Ptrain}\left[w(X)R(X)\right] - \alphaP  \nonumber\\*
{\rm subject~to} \qquad & w_{\myhash} \Ptrain \in \calB, \quad w \in \Cisow.
\end{align}
For comparison, we also consider the following optimization problem
\begin{align}\label{eq:wt-delta-tilde}
\tDisow \quad = \quad \sup_{h:~h \circ \wprior \geq 0} \qquad & \EE_{\Ptrain}\left[(h \circ \wprior)(X)R(X)\right] - \alphaP  \nonumber\\*
{\rm subject~to} \qquad & \left(h \circ \wprior\right)_{\myhash} \Ptrain \in \calB, \quad h \in \Cisor,
\end{align}
where $\Cisor$ denotes the cone of isotonic functions defined on $\RR$ equipped with the natural ordering.
In fact, since $\Cisow = \{h\circ w_0 : h\in\Cisor\}$ by definition, we therefore have $\Disow = \tDisow$.

In addition, recalling that $\tilde R(\wprior(X)) = \EE_{\Ptrain}[R(X)\mid \wprior(X)]$, by the change of measure, the optimization problem in \eqref{eq:wt-delta-tilde} can be further rewritten as 
\begin{align}\label{eq:wt-delta-equiv}
\Disow \quad = \quad \sup_{h:~h \geq 0} \qquad & \EE_{(\wprior)_{\myhash}\Ptrain}\left[h(U)\tilde R(U)\right] - \EE_{(\wprior)_{\myhash}\Ptrain}\left[\tilde R(U)\right] \nonumber\\*
{\rm subject~to} \qquad & (h \circ \wprior)_{\myhash}\Ptrain \in \calB, \quad h \in \Cisor.
\end{align}
We observe that \eqref{eq:wt-delta-equiv} has the same form with the definition of $\Diso$ in \eqref{eq:opt-iso0}, where we consider the probability measure $(\wprior)_{\myhash} \Ptrain$ instead of $\Ptrain$  and $\tilde R$ in place of $R$, and with the specific isotonic cone $\Cisor$ on $\RR$.

Applying Theorem~\ref{thm:equiv} to \eqref{eq:wt-delta-equiv} yields
\begin{align}\label{eq:wtd-delta-proj}
\Disow \quad = \quad \sup_{h:~h \geq 0} \qquad & \EE_{(\wprior)_{\myhash}\Ptrain}\left[h(U)[\pi_1(\tilde R)](U)\right] - \EE_{(\wprior)_{\myhash}\Ptrain}\left[\tilde R(U)\right] \nonumber\\*
{\rm subject~to} \qquad & (h \circ \wprior)_{\myhash}\Ptrain \in \calB,
\end{align}
where $\pi_1$ is the projection onto $\Cisor$ under the measure $(\wprior)_{\myhash}\Ptrain$. 
By definition of $\tilde R$, we can rewrite this as
\begin{align*}
\Disow \quad = \quad \sup_{h:~h \geq 0} \qquad & \EE_{\Ptrain}\left[h(w_0(X))[\pi_1( \tilde R)](w_0(X))\right] - \EE_{\Ptrain}\left[\tilde R(w_0(X))\right] \nonumber\\*
{\rm subject~to} \qquad & (h \circ \wprior)_{\myhash}\Ptrain \in \calB,
\end{align*}
which is equal to $\altDwiso$ as defined in \eqref{eq:wt-recalib} since we also have $ \EE_{\Ptrain}\left[\tilde R(w_0(X))\right] =  \EE_{\Ptrain}\left[[\pi_1(\tilde R)](w_0(X))\right]$
by~\eqref{eqn:iso_fact_1}.
We herein complete the proof.

\subsection{A misspecified isotonic constraint}\label{sec:mis-iso}
When the true distribution shift does not obey the isotonic constraint exactly, we can nonetheless 
provide a bound on the worst-case excess risk, which is 
tighter than the (non-iso) DRL bound whenever the isotonic constraint provides a reasonable approximation. 

Denote $\tilde w^*$ as the underlying density ratio $\mathsf{d}\Ptar/\mathsf{d}\Ptrain$ and $\Delta^*(R) = \EE_{\Ptrain}[\tilde w^*(X)R(X)] - \EE_{\Ptrain}[R(X)]$ as the true excess risk. 
Then, we have the following connections between $\Delta^*(R)$ and $\Diso$.
\begin{proposition}\label{thm:violation-lower-bnd}
Assume Condition~\ref{cond:B-cond} holds. If $\tilde w^*_{\myhash} \Ptrain \in \calB$ and $\tilde w^*\in L_2(\Ptrain)$, then we have 
\[
\Delta^*(R)\leq \Diso + \EE_{\Ptrain}\bigg[[\tilde w^* - \pi(\tilde w^*)](X) \cdot [R - \pi(R)](X)\bigg].
\] 
In particular, if either $\tilde w^*\in\Ciso$ or $R \in \Ciso$, then $ \Delta^*(R) \leq \Diso$.
\end{proposition}

The result states that when the isotonic constraint is violated, the worst-case excess risk of iso-DRL will be no worse than the true excess risk plus a gap which can be controlled by the correlation between $[\tilde w^* - \pi(\tilde w^*)](X)$ and $[R - \pi(R)](X)$. In particular, if \emph{either} the risk or the true density ratio is itself isotonic
(or approximately isotonic), then the gap term must be zero (or approximately zero)---and so the excess risk calculation $\Diso$, which is tighter than the non-iso DRL bound $\Du$, will never underestimate the true risk $\Delta^*(R)$ (or will only
be a mild underestimate).

\subsubsection{Proof of Proposition~\ref{thm:violation-lower-bnd}}\label{sec:proof_thm:violation-lower-bnd}

Recall that $\tilde w^*$ is the underlying density ratio $\mathsf{d}\Ptar/\mathsf{d}\Ptrain$.
Since $\pi(\tilde w^*) \overset{cvx}{\orderless} w^*$ by~\eqref{eqn:iso_fact_4}, and $\calB$ is closed under the convex ordering by Condition~\ref{cond:B-cond}, we have $\pi(\tilde w^*)_{\myhash} \Ptrain \in \calB$; of course, we also have  $\pi(\tilde w^*) \in \Ciso$  by definition.
Therefore, $\pi(\tilde w^*)$ is feasible for the optimization problem~\eqref{eq:opt-iso0}, and so we have \[\Diso \geq \EE_{\Ptrain}\bigg[[\pi(\tilde w^*)](X)R(X)\bigg] - \alphaP.\]
We therefore have
\[\Delta^*(R) = \EE_{\Ptrain}[\tilde w^*(X)R(X)] - \EE_{\Ptrain}[R(X)] \\\leq  \Diso +\EE_{\Ptrain}\left[\Big(\tilde w^*(X) - [\pi(\tilde w^*)](X)\Big)R(X)\right].\]
Moreover,
\[\EE_{\Ptrain}\left[\Big(\tilde w^*(X) - [\pi(\tilde w^*)](X)\Big)\cdot [\pi(R)](X)\right]
= \langle \pi(R), \tilde w^* - \pi(\tilde w^*)\rangle_{\Ptrain} \leq 0\]
by~\eqref{eqn:iso_fact_2}, and so we have
\[\EE_{\Ptrain}\left[\Big(\tilde w^*(X) - [\pi(\tilde w^*)](X)\Big)R(X)\right]
\leq \EE_{\Ptrain}\left[\Big(\tilde w^*(X) - [\pi(\tilde w^*)](X)\Big)\cdot \big(R(X) - [\pi(R)](X)\big)\right].\]
This completes the proof.

\subsection{Proof of Proposition~\ref{prop:l2-approx}}\label{sec:proof-l2}

For any $w \geq 0$, we define the sequence of truncated functions $\{w_n\}_{n \in \NN}$ via
\[
w_n(x) = w(x) \cdot \ind\{w(x) \leq n\} + L_n \cdot \ind\{w(x) > n\},
\]
where $L_n = \EE[w(X) \mid w(X) > n]$. By construction for each $n$, $\EE_{\Ptrain}[w_n(X)] = 1$ and,
since $\max \{n, L_n\} = L_n < \infty$,  $w_n\in L_2(\Ptrain)$ for each $n\geq 1$. 

\paragraph{Step 1: Feasibility of $w_n$.} We first prove the feasibility of $w_n$. To see this, as $\EE_{\Ptrain}[w_n(X)] = 1$ by construction, we need to show that $(w_n)_{\myhash} \Ptrain \in \calB$. By Condition~\ref{cond:B-cond}, since $\calB$ is closed under the convex ordering, it suffices to show that
\[
\EE_{\Ptrain}\left[\psi(w_n(X))\right] \leq \EE_{\Ptrain}\left[\psi(w(X))\right] \qquad \textnormal{ for any convex function }\psi.
\]
This is true by Jensen's inequality, since, by construction, $\EE_{\Ptrain}[w(X)\mid w_n(X)] = w_n(X)$.

\paragraph{Step 2: Convergence of $\EE_{\Ptrain}[w_n(X)R(X)]$.} To verify the convergence of $\EE_{\Ptrain}[w_n(X)R(X)]$, consider
\begin{align*}
& \bigg| \EE_{\Ptrain}[w_n(X)R(X)] - \EE_{\Ptrain}[w(X)R(X)] \bigg|\\* 
=\;& \bigg|\int_{w(x) > n} (L_n - w(x)) R(x) \mathsf{d}\Ptrain(x)\bigg|\\*
\leq\; & B_R \int_{w(x) > n} \big| L_n - w(x) \big| \mathsf{d}\Ptrain(x)\\*
\leq\; & B_R \left(\int_{w(x) > n} w(x) \mathsf{d}\Ptrain(x) + L_n \PP(w(X) > n)\right)\\*
=\; & 2 \EE_{\Ptrain}\left[w(X) \cdot \ind\{w(X) > n\}\right].
\end{align*}
Finally, since $\EE_{\Ptrain}[w(X)] = 1$ (i.e., we know that $w\in L_1(\Ptrain)$), this means that \[\lim_{n\to \infty} \EE_{\Ptrain}\left[w(X) \cdot \ind\{w(X) > n\}\right] =0.\]

\paragraph{Conclusion.}
For any $\vareps > 0$, there exists $w \geq 0$ such that $\EE_{\Ptrain}[w(X)] = 1$, $w_{\myhash}\Ptrain \in \calB$, and 
\[
\EE_{\Ptrain}[w(X)R(X)] - \EE_{\Ptrain}[R(X)] \geq \Du - \vareps/2
\]
Then, based on Step 2, for
sufficiently large $n$ it holds that $\EE_{\Ptrain}[w_n(X)R(X)] \geq \EE_{\Ptrain}[w(X)R(X)] - \vareps/2$.
From Step 1, we know that $w_n$ is feasible for $\Dul$, i.e.,
\[\Dul \geq \EE_{\Ptrain}[w_n(X)R(X)] - \EE_{\Ptrain}[R(X)] \geq  \left( \EE_{\Ptrain}[w(X)R(X)] - \vareps/2\right) - \EE_{\Ptrain}[R(X)] 
\geq \Du - \vareps.\]
Since $\vareps$ is arbitrary this verifies that $\Dul \geq \Du$, and clearly we must have $\Dul \leq \Du$ by construction, which completes the proof.

\section{Proofs  of results in Section~\ref{sec:conv}}\label{sec:proof-conv}

\subsection{Proof of Proposition~\ref{thm:thirdterm_2}}
To prove the proposition, it suffices to show that $\|\wfiso\|_{\infty} < \infty$.
Recall the dual formulation. There exists a pair $(\lambda^*, \nu^*)$ such that
\[
\wfiso(x) = \calP_{[0,+\infty)}\left\{(f^\prime)^{-1}\left(\frac{[\pi(R)](x) - \nu^*}{\lambda^*}\right)\right\}.
\]
Note that $\nu^*$ is the parameter for standardization, thus to guarantee $\EE_{\Ptrain}[\wfiso(X)] = 1$, we have
\[
(f^\prime)^{-1}\left(\frac{B_R - \nu^*}{\lambda^*}\right) \geq \sup_{x \in \calX}\; \wfiso(x) \geq 1.
\]
Moreover, it holds that $(f^\prime)^{-1}(-\nu^*/\lambda^*) \leq \min_{x \in \calX}\; \wfiso(x) \leq 1$. Then, combining the inequalities yields 
\begin{align}\label{ineq:nu}
-\lambda^* f^\prime(1) \leq \nu^* \leq B_R - \lambda^* f^\prime(1).
\end{align}
If we further have $\lambda^* \geq \underline{\lambda} > 0$, it holds that 
\[
\|\wfiso\|_{\infty} \leq (f^{\prime})^{-1}\left(\frac{B_R + \lambda^* f^\prime(1)}{\lambda^*}\right) \leq (f^{\prime})^{-1}\left(f^\prime(1) + \frac{B_R}{\underline{\lambda}}\right) < \infty.
\]
Then, it remains to prove that $\lambda^* \neq 0$. To see this, consider the KKT condition:
\begin{align*}
-[\pi(R)](x) + \lambda^* f^\prime(\wfiso(x)) + \nu^* & = 0,\\*
\lambda^* \left(\EE_{\Ptrain}[f(\wfiso(X))] - \rho\right) &= 0,\\*
\nu^* \left(\EE_{\Ptrain}[\wfiso(X)] - 1\right) &= 0.
\end{align*}
If $\lambda^* = 0$, we have $[\pi(R)](X) = \nu^*$ $\Ptrain$-almost surely, which implies that $\wfiso(X) = 1$ $\Ptrain$-almost surely, in which case $\wfiso$ is also bounded. Combining pieces above, we have shown that $\|\wfiso\|_{\infty} < \infty$.

\subsection{Proof of Theorem~\ref{thm:deterministic}}

We first fix any $w\in\Ciso$ with $w_{\myhash}\Ptrain\in\calB$. By Condition~\ref{cond:bnd-w}, it holds that $w(X)\leq \wbnd$ $\Ptrain$-almost surely,
and therefore without loss of generality we can assume $w\in\CisoB$. Then, by definition of $\vareps_{\calB}$, for any $\delta>0$, we can find some $s,t\geq 0$ 
with $s+t \leq \vareps_{\calB} + \delta$, such that we have $w'_{\myhash}\hat P_n \in\calB$ by defining $w' = (1-s)\cdot w + t\cdot\mathbf{1}$.

Moreover, by construction, we must have $w'\in\Ciso$. Therefore, by optimality, we have
\begin{align*}
\hatDisoU & \geq \frac{1}{n}\sum_{i=1}^n w'(X_i)r(X_i,Y_i) - \frac{1}{n}\sum_{i=1}^nr(X_i,Y_i)\\* &= \EE_{\hat P_n}[(w'(X)-1)r(X,Y)]\\* &\geq
\EE_{\Ptrain}[(w'(X)-1)R(X)]  - \vareps_R,
\end{align*}
where the last inequality is by the definition of $\vareps_R$. Plugging in the definition of $w'$, we obtain that
\begin{align*}\hatDisoU & \geq \EE_{\Ptrain}[\left((1-s)w(X) + t)-1\right)R(X)]  - \vareps_R\\*
& = (1-s)\EE_{\Ptrain}[(w(X)-1)R(X)] + (t-s)\EE_{\Ptrain}[R(X)] - \vareps_R\\* 
& = \EE_{\Ptrain}[(w(X)-1)R(X)] - (s+t)\EE_{\Ptrain}[w(X)R(X)] + t\EE_{\Ptrain}[(w(X) + 1)R(X)] - \vareps_R\\* 
&\geq \EE_{\Ptrain}[(w(X)-1)R(X)] -2 B_R\wbnd \cdot \vareps_{\calB}-  \vareps_R ,\end{align*}
where the last inequality is by the fact that $\|w\|_\infty\leq \wbnd$ and $R$ is $B_R$-bounded, and $\wbnd\geq 1$. Since this holds for every $w\in\Ciso$ with $w_{\myhash}\Ptrain\in\calB$, by definition of $\Diso$, we therefore have
\[\hatDisoU \geq \Diso -  \vareps_R - 2B_R\wbnd \cdot \vareps_{\calB}. \]

By identical arguments, with the roles of $\Ptrain$ and $\hat P_n$ reversed, we can also show that
\[\Diso \geq \hatDisoU -  \vareps_R - 2B_R\wbnd \cdot \vareps_{\calB},\]
which completes the proof.

\subsection{Proof of Lemma~\ref{lem:vareps_R}}
Throughout this proof we will use the notation of supervised learning, since unsupervised learning can be viewed as a special case.

In the first step, we will bound $\EE[\vareps_R]$. By symmetrization  (\cite{wellner2013weak} Theorem 2.3.1), we have
\begin{align*}\EE[\vareps_R] &= \EE\left[\sup_{w\in\CisoB} \left|\EE_{\hat P_n}[(w(X)-1)r(X,Y)] - \EE_{\Ptrain}[(w(X)-1)R(X)]\right|\right]\\
&\leq 2\EE\left[\sup_{w\in\CisoB} \left|\frac{1}{n}\sum_{i=1}^n \sigma_i (w(X_i)-1)r(X_i,Y_i)\right|\right],\end{align*}
where $\sigma_i$'s are independent Unif$\{\pm 1\}$ random variables.
Since risk is $B_R$-bounded, by the Ledoux-Talagrand contraction lemma (\cite{ledoux2013probability} Theorem 4.12) applied
with functions $\phi_i(t) = (t-1)\cdot r(X_i,Y_i)$, we further have
\[\EE\left[\sup_{w\in\CisoB} \left|\frac{1}{n}\sum_{i=1}^n \sigma_i (w(X_i)-1)r(X_i,Y_i)\right|\right]
\leq 2B_R \EE\left[\sup_{w\in\CisoB} \left|\frac{1}{n}\sum_{i=1}^n \sigma_i w(X_i)\right|\right] = 2B_R\calR_n(\CisoB).\]

Now we bound $\vareps_R$ with high probability.  Since risk is $B_R$-bounded, and any function $w\in\CisoB$ is $\wbnd$-bounded, 
we have $(w(X)-1)r(X,Y) \in [-B_R,(\wbnd-1)B_R]$, and so resampling one data point can perturb $\vareps_R$ by at most $\wbnd B_R/n$.
Therefore, by McDiarmid's inequality  \citep{mcdiarmid1989method}, with probability at least $1-n^{-1}$, it holds that
\[\vareps_R \leq \EE[\vareps_R] +  B_R\wbnd \sqrt{\frac{\log n}{2n}}.\]
Combining all these calculations yields the desired bound.

\subsection{Proof of Lemma~\ref{lem:vareps_calB_a_b}}
Recall that
\[\vareps_{\calB} = \sup_{w\in\CisoB} \max\left\{ \vareps_{\calB}\left(w; \Ptrain,\hat P_n\right),  \vareps_{\calB}\left(w; \hat P_n,\Ptrain\right)\right\}, \]
where
\[\vareps_{\calB} \left(w ; P_0, P_1\right)  =\inf\left\{ s\geq  0 \ : \  \exists \ t\geq 0, \, \big((1-s)\cdot w + t\cdot \mathbf{1}\big)_{\myhash}P_1\in\calB\right\}.\]
First, following the exact same steps as in the proof of Lemma~\ref{lem:vareps_R}, with the notation $\delta_w =  \EE_{\Ptrain}[w(X)] - \EE_{\hat P_n}[w(X)]$, we have
\begin{align}\label{eq:event-1}
\sup_{w\in\CisoB} \left|\delta_w\right|\leq 4\calR_n(\CisoB) + \wbnd \sqrt{\frac{\log n}{2n}} =: \vareps'
\end{align}
with probability at least $1-n^{-1}$. 

Assume the event~\eqref{eq:event-1} holds.
Fix any $w\in\CisoB$ with $w_\myhash\Ptrain\in\calB_{a,b}$, and define
\[w' = (1-s)\cdot w + t \cdot \mathbf{1},\]
where $s, t \geq 0$ are chosen such that $\EE_{\hat P_n}[w'(X)] = 1$, indicating that $t = s + (1-s)\delta_w$.

If $\vareps' = 4\calR_n(\CisoB) + \wbnd \sqrt{\frac{\log n}{2n}} > \frac{1}{2}\min\left\{1-a, b-1\right\}$,
then since $\vareps_{\calB}\leq 1$ holds by definition, the result of the lemma must hold trivially. Therefore we can 
restrict our attention to the case that
\[\vareps' \leq  \frac{1}{2}\min\left\{1-a, b-1\right\}.\]
We can further choose  
\[
s = 2\max\left\{\frac{\vareps'}{b-1}, \frac{\vareps'}{1-a}\right\} \geq \max\left\{\frac{\vareps'}{b-1-\vareps'}, \frac{\vareps'}{1-a-\vareps'}\right\},
\]
with which, we can verify that
\[
w'(X) \leq (1-s) b + t = (1-s)(b + \delta_w) + s \leq (b+\vareps') + s(1-b+\vareps') \leq b,
\]
and similarly, $w'(X) \geq a$.
Therefore, we have $w'_\myhash \hat P_n\in\calB_{a,b}$.

The same construction holds with the roles of $\Ptrain$ and $\hat P_n$ reversed. Therefore, we can take $\vareps_{\calB}=s$, which completes the proof.

\subsection{Proof of Lemma~\ref{lem:vareps_calB_f_rho}}
First, following the same steps (i.e., symmetrization and contraction) as in the proof of Lemma~\ref{lem:vareps_R}, we have
\begin{align}\label{eq:event-2}
\sup_{w\in\CisoB} \left|\EE_{\hat P_n}[w(X)] - \EE_{\Ptrain}[w(X)]\right|\leq 4\calR_n(\CisoB) + \wbnd \sqrt{\frac{\log n}{2n}} =: \vareps'
\end{align}
with probability at least $1-n^{-1}$. 

Moreover, denote $t^*_f = \argmin {t \in [0,\wbnd]} f(t)$.
We have the decomposition \[f(t) = f(t) \cdot \ind\{f(t) \geq t^*_f\} + f(t) \cdot \ind\{f(t) < t^*_f\} \eqqcolon f_1 + f_2,\] where both $f_1$ and $-f_2$ are nondecreasing. Then, for any $g = f \circ w$ with $w \in \CisoB$, we have the decomposition $g = f_1 \circ w + f_2 \circ w$, where $f_1 \circ w \in \Ciso$, $-f_2 \circ w \in \Ciso$, and both functions $f_1$, $f_2$ are $L_{\wbnd}$-Lipschitz. Then, by the Ledoux-Talagrand contraction lemma (\cite{ledoux2013probability} Theorem 4.12) applied
with functions $\phi_i(t) = f(t)$, we have
\begin{align*}
\calR_n\left(\left\{g = f \circ w:\;w\in \CisoB\right\}\right) & \leq 2\calR_n\left(\left\{g = f \circ w:\;w\in \CisoB,\;f\textnormal{ is nondecreasing and }L_{\wbnd}\textnormal{-Lipschitz}\right\}\right)\\
&\leq 8L_{\wbnd}\calR_n(\CisoB).
\end{align*}
Hence, similar to the proof of Lemma~\ref{lem:vareps_R}, we have
\begin{align}\label{eq:event-3}
\sup_{w\in\CisoB} \left|\EE_{\hat P_n}[f(w(X))] - \EE_{\Ptrain}[f(w(X))]\right|\leq 8L_{\wbnd}\calR_n(\CisoB) + L_{\wbnd} \wbnd \sqrt{\frac{\log n}{2n}} =: \vareps''
\end{align}
with probability at least $1-n^{-1}$.

Assume events~\eqref{eq:event-2} and \eqref{eq:event-3} both hold.
Fix any $w\in\CisoB$ with $w_\myhash\Ptrain\in\calB_{f,\rho}$, and define 
\[w' = (1-s)\cdot w + t \cdot \mathbf{1},\]
where $s, t \in (0,1)$ are chosen such that $\EE_{\hat P_n}[w'(X)] = 1$, which implies that $t = s + (1-s)\delta_w$.

Since $f$ is $L_{\wbnd}$-Lipschitz on $[0,\wbnd]$,
\[f(w'(x)) \leq f\left((1-s)\cdot  w(x) + s\right) + L_{\wbnd}  \cdot |t-s| \leq f\left((1-s)\cdot  w(x) + s\right) + L_{\wbnd}  (1-s)|\delta_w|.\] 
And, since $f$ is convex with $f(1)=0$,
\[ f\left((1-s)\cdot  w(x) + s\right) \leq (1-s) f(w(x)) + s f(1) = (1-s) f(w(x)).\]
Combining everything, for all $x$, it holds that
\[f(w'(x))\leq  (1-s) f(w(x))+  L_{\wbnd}  (1-s)|\delta_w| \leq (1-s) \left(f(w(x)) + L_{\wbnd} \vareps'\right).\] 
Hence, we have
\[\EE_{\hat P_n}[ f(w'(X))]  \leq (1-s) \EE_{\hat P_n}[ f(w(X))]+  (1-s)L_{\wbnd} \vareps' .\]
And by assumption, $ \EE_{\hat P_n}[ f(w(X))]\leq  \EE_{\Ptrain}[ f(w(X))] + \vareps'' \leq \rho +\vareps''$, so,
\[\EE_{\hat P_n}[ f(w'(X))]  \leq (1-s) \cdot(\rho +\vareps'' + L_{\wbnd} \vareps')\leq\rho,\]
where the last step holds by choosing
\[
s = \frac{1}{\rho}(\vareps'' + L_{\wbnd} \vareps') \geq \frac{\vareps'' + L_{\wbnd} \vareps'}{\rho +\vareps'' + L_{\wbnd} \vareps'}.
\]
This verifies that $w'_{\myhash}\hat P_n\in\calB_{f,\rho}$.

The same construction holds with the roles of $\Ptrain$ and $\hat P_n$ reversed. Therefore, we can take $\vareps_{\calB}=s$, which completes the proof.

\subsection{The role of the isotonic constraint}\label{sec:iso-role}

The consistency bounds developed above show that, under appropriate conditions,
the error in estimating $\Diso$ can be controlled whenever the appropriate Rademacher complexity terms are small. 
This suggests that the isotonic constraint plays an important role: essentially, the isotonic constraint induces a form of regularization, ensuring that we work with a low-complexity class of functions.

To verify this, we now
present 
an example with the constraint set $\calB=\calB_{a,b}$, \emph{without} an isotonic constraint,
where the estimation error of the (non-iso) DRL risk does not converge to zero.

To make the question more concrete, we will work with the bound constraint $\calB_{a,b}$ with $0 \leq a \leq 1 \leq b$,
and consider the optimization problem
\begin{align*}
\hatDbndr \quad = \quad \max_{w \geq 0} \qquad  \frac{1}{n} \sum_{i \leq n} w(X_i) r_i - \frac{1}{n}\sum_{i \leq n} r_i  \qquad {\rm subject~to} \quad w_{\myhash} \hat P_n \in \calB_{a,b},
\end{align*}
which estimates the excess risk without the isotonic constraint.
In other words, using $\hatDbndr$ as an empirical estimate of $\Dbnd$, is analogous to using $\hatDbndisor$ as an empirical estimate of $\Dbndiso$ in the 
presence of an additional isotonic constraint.

The following result shows that, without an isotonic constraint, this empirical estimate is \emph{not} a consistent estimator
of the true excess risk. 
\begin{proposition}\label{prop:noniso-lower}
Assume $R(X) = 1/2$ holds $\Ptrain$-almost surely. Then, $\Dbnd = 0$,
but with probability at least $1 - 2e^{-n/24}$, it holds that $\hatDbndr  \geq \min\{1-a,b-1\}/16$.
\end{proposition}
In other words, $\hatDbndr$ is not a consistent estimator of the true excess risk $\Dbnd$, since the error in the estimate
is bounded away from zero (as long as $a<1<b$). This means that the constraint set $\calB_{a,b}$, on its own,
is not sufficiently constrained to enable consistent estimation---while in contrast, as we have seen in our theoretical guarantees
for estimation for iso-DRL, adding an isotonic constraint enables the excess risk to be estimated consistently with an empirical sample.

\subsubsection{Proof of Proposition~\ref{prop:noniso-lower}}\label{sec:hardness}

By construction, we have $r_i \sim {\rm Bern}(R(X_i)) = {\rm Bern}(1/2)$ independently for $i=1,\dots,n$. According to Section~\ref{sec:sa_bnd}, the worst-case weights take the form $w_i = w(X_i) = c_1 \cdot \ind\{r_i = 0\} + c_2 \cdot \ind\{r_i = 1\}$, where $a \leq c_1 \leq 1 \leq c_2 \leq b$. Moreover, by the KKT condition, at least one of $c_1 = a$ and $c_2 = b$ holds, which implies that $c_2 - c_1 \geq \min\{1-a, b-1\} \eqqcolon \delta$.
Then, the estimated excess risk can be expressed as
\begin{align*}
\hatDbndr & = \frac{c_2}{n} \sum_{i \leq n} r_i - \frac{1}{n} \sum_{i \leq n} r_i = \frac{c_2-1}{n} \sum_{i \leq n} r_i.
\end{align*}
Since $n^{-1}\sum_{i \leq n}w_i = 1$, we have
\[
\frac{1}{n}\sum_{i \leq n} (1-r_i) = \frac{c_2-1}{c_2 - c_1},
\]
which implies 
\[
c_2 - 1 = \frac{c_2 - c_1}{n}\sum_{i \leq n} (1-r_i) \geq \frac{\delta}{n}\sum_{i \leq n} \left(1-r_i\right).
\]

In the meantime, by Chernoff bounds, with probability at least $1 - 2e^{-n/24}$, it holds that
\[
\bigg|\frac{1}{n} \sum_{i \leq n}r_i - \frac{1}{2}\bigg| \leq \frac{1}{4}.
\]
Then, for the excess risk, with probability at least $1 - 2e^{-n/24}$, it holds that
\begin{align*}
\hatDbndr = \frac{c_2-1}{n} \sum_{i \leq n} r_i \geq \delta \left(\frac{1}{n}\sum_{i \leq n} \left(1-r_i\right)\right) \cdot \left(\frac{1}{n} \sum_{i \leq n} r_i\right) \geq \frac{\delta}{16}.
\end{align*}

\section{Additional simulation results}\label{sec:add-simu}

\subsection{Simulations for iso-DRL under componentwise order}\label{sec:add-simu_preorder}
In Section~\ref{sec:simu}, we mainly focused on the partial order with respect to $\wprior(x)$. In this section, to demonstrate the effect of various choices of the partial (pre)order, we further consider an alternative choice of the partial (pre)order: the componentwise order where
\[
x \orderless x^{\prime} \qquad \textnormal{if and only if} \qquad x_j \leq x^{\prime}_j,\;\textnormal{for all }j\in [m],
\]
where we set $m = 5 < d = 20$.
Let iso-DRL-comp denote the CP interval with calibrated target level $\alpha^{\prime}_{\iso} = \max\{0, \alpha - \tilde \Delta^{\iso}\}$, where
\begin{align}\label{eq:delta-hat-isoDRL-comp}
\tilde \Delta^{\iso} = &\max \qquad  \frac{1}{n} \sum_{i \in \calD_3} w_i \tilde r^{\iso}_i - \frac{1}{n} \sum_{i \in \calD_3} r_i\nonumber\\* &{\rm subject~to} \qquad \frac{1}{n}\sum_{i \in \calD_3} w_i=1, \quad \frac{1}{n}\sum_{i \in \calD_3} w_i \log w_i \leq \rho, \quad 0 \leq w_i \leq \wbnd,
\end{align}
and $(\tilde r_i)_{i \in \calD_3}$ is the isotonic projection of $(r_i)_{i \in \calD_3}$ with respect to the componentwise order.

\begin{figure}[htb]
  \centering 
    \includegraphics[width=0.68\textwidth]{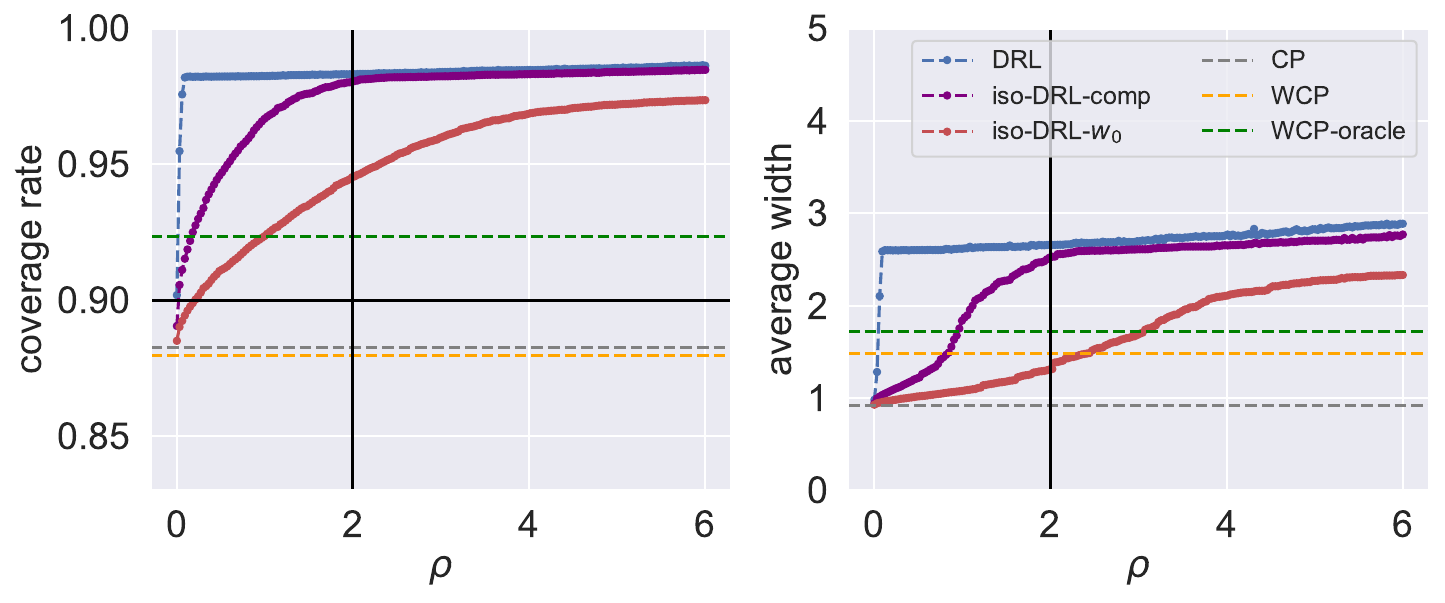}
    \caption{Results with varying $\rho$ in the well-specified setting. The solid vertical line denotes an estimate $\widehat\rho$ of
    the KL divergence, $D_{\rm KL}(\Ptar \| \Ptrain)$ (See Appendix~\ref{sec:add-simu_wine} for details).
    The solid horizontal line (in the left-hand plot) marks the nominal coverage level, $1-\alpha = 90\%$.}
    \label{fig:generic_1_}
\end{figure}

\begin{figure}[htb]
  \centering 
    \includegraphics[width=0.68\textwidth]{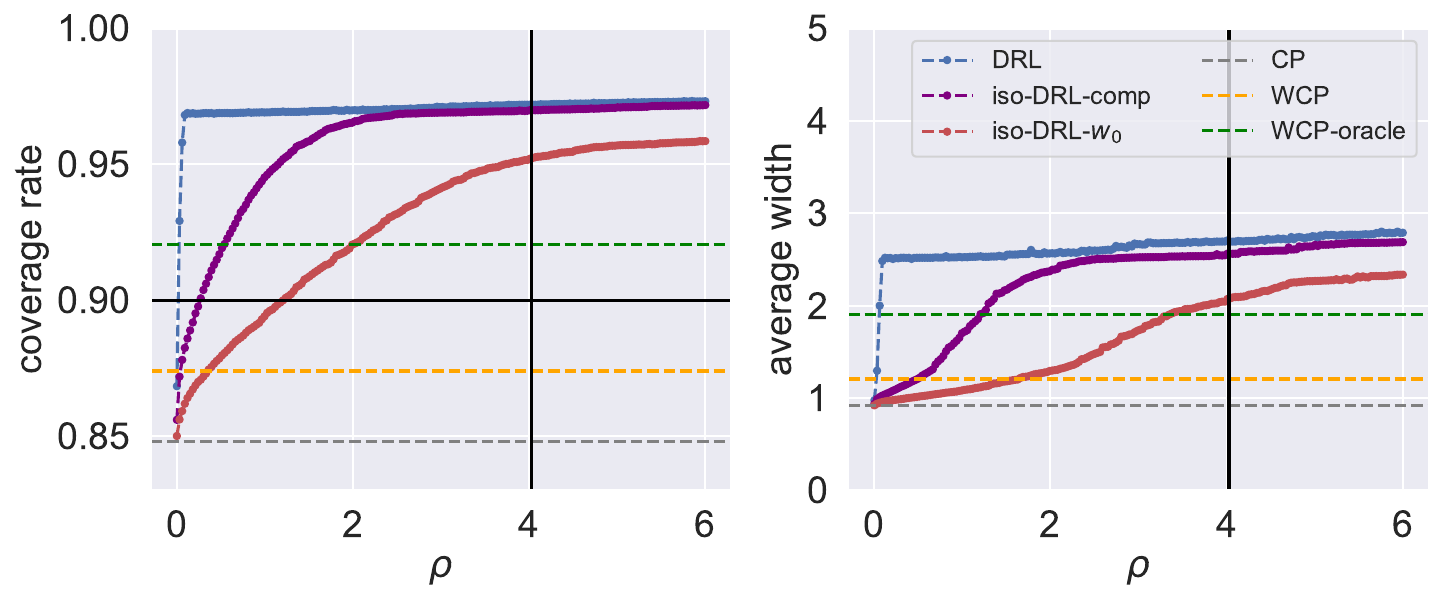}
    \caption{Results with varying $\rho$ in the misspecified setting. The solid vertical line denotes an estimate $\widehat\rho$ of the KL divergence, $D_{\rm KL}(\Ptar \| \Ptrain)$ (See Appendix~\ref{sec:add-simu_wine} for details).
     The solid horizontal line (in the left-hand plot) marks the nominal coverage level, $1-\alpha = 90\%$.} 
    \label{fig:generic_2_}
\end{figure}
We follow exactly the same settings with Section~\ref{sec:synthetic} with $n_{\rm pre} = 50$ and vary $\rho$ in $[0.002, 6]$. From Figure~\ref{fig:generic_1_} and \ref{fig:generic_2_}, each of the coverage rate and average interval width of iso-DRL-comp lies between that of DRL and iso-DRL-$\wprior$, which indicates that additional constraints will relieve the conservativeness of DRL, but only a proper choice of the partial (pre)order will lead to desired performance close to the oracle weighted CP.

\subsection{Details for the \texttt{wine quality} data set: a proxy of the oracle KL-divergence}\label{sec:add-simu_wine}
In this section, we examine the choice of $\rho$ in the \texttt{wine quality} data experiment from Section~\ref{sec:real-simu}.
In a real data setting, the true KL divergence, $D_{\rm KL}\left(\Ptar\|\Ptrain\right)$, is of course unknown, so we need
to use a data-driven choice of $\rho$ in order to implement a DRL procedure (with or without an isotonic constraint).

As is shown in Section~\ref{sec:real-simu}, we denote $\hat w_{\textnormal{kde}}$ as the density ratio obtained by kernel density estimation (Gaussian kernel with bandwidth $0.125$). Accordingly, let $\mathsf{d}\hat Q_{\textnormal{kde}} = \hat w_{\textnormal{kde}} \cdot \mathsf{d}P$ be an estimate of $\Ptar$. With a subsample $\{X_i\}_{i \leq K}$ drawn the group of white wine (data distribution $\Ptrain$), a reasonable value for $\hat \rho$ (i.e., an estimate of the true divergence $\rho$ between the distributions $\Ptrain$ and $\Ptar$) can be calculated by
\begin{align*}
\hat \rho & = \frac{1}{K} \sum_{i \leq K} \hat w_{\textnormal{kde}}(X_i) \log\left(\hat w_{\textnormal{kde}}(X_i)\right)\\* & \approx \EE_{\Ptrain}\left\{\frac{\mathsf{d}\hat Q_{\textnormal{kde}}}{\mathsf{d}\Ptrain} \log\left(\frac{\mathsf{d}\hat Q_{\textnormal{kde}}}{ \mathsf{d}\Ptrain}\right)\right\} = D_{\rm KL}\left(\hat Q_{\textnormal{kde}} \| \Ptrain\right).
\end{align*}
\begin{figure}[htb]   
    \centering 
    \includegraphics[width=0.45\textwidth]{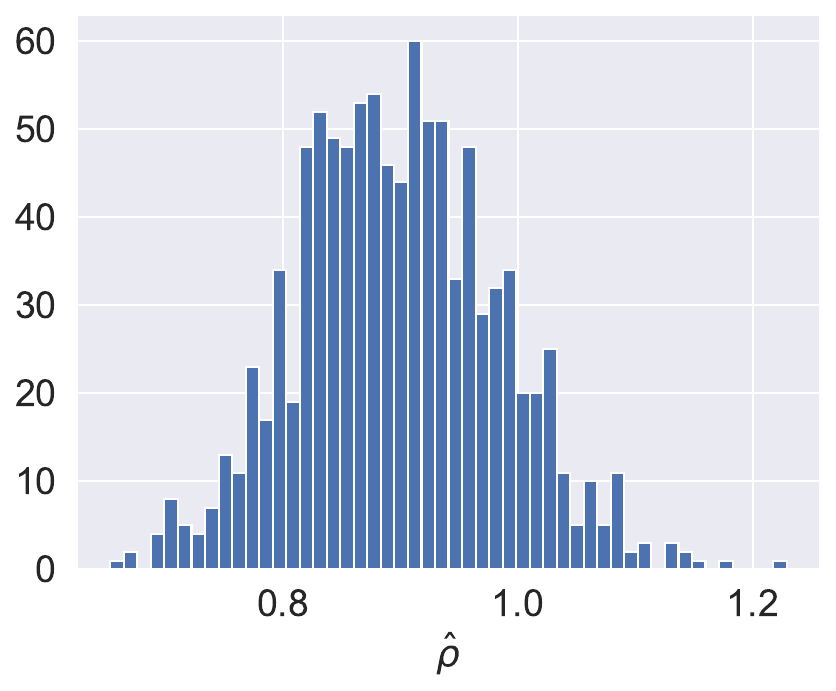}
    \caption{Histogram of $\hat \rho$. (See Appendix~\ref{sec:add-simu_wine} for details.)} 
    \label{fig:wine-rho}
\end{figure}

To show the range for values of $\hat \rho$, we repeatedly fit KDE on the $80\%$ samples from each group (white and red wine groups respectively). 
Figure~\ref{fig:wine-rho} shows the histogram of $\hat \rho$ with $1000$ repetitions, of which the median is approximately $0.8950$---this 
is the value of $\rho$ used in our preview of the \texttt{wine quality} data experiment, shown in Figure~\ref{fig:eg1-2}.

\bibliography{main}
\bibliographystyle{apalike}
\end{document}